\documentclass[aps,pre,twocolumn,groupedaddress,superscriptaddress,showpacs,preprintnumbers,amsmath,amssymb]{revtex4-1}

 \usepackage{graphicx}
 \usepackage{dcolumn}
 \usepackage{bm}

\newcommand{\be}{\begin{equation}} \newcommand{\ee}{\end{equation}}
\newcommand{\bea}{\begin{eqnarray}} \newcommand{\eea}{\end{eqnarray}}

\begin{document}

\title{Percolation transitions in the survival of interdependent agents on multiplex networks, 
              catastrophic cascades, and SOS}

\author{Peter Grassberger}\affiliation{JSC, FZ J\"ulich, D-52425 J\"ulich, Germany}
                          \affiliation{Institute for Advanced Studies in Basic Sciences, Gava Zang, 
                           Zanjan 45137-66731, Iran}

\date{\today}

\begin{abstract}
The ``SOS" in the title does not refer to the international distress signal, but to ``solid-on-solid" (SOS)
surface growth. The catastrophic cascades are those observed by Buldyrev {\it et al.} in interdependent 
networks, which we re-interpret as multiplex networks with agents that can only survive if 
they mutually support each other, and whose survival struggle
we map onto an SOS type growth model. This mapping not only reveals non-trivial structures in the phase space 
of the model, but also leads to a new and extremely efficient simulation algorithm. We use
this algorithm to study interdependent agents on duplex Erd\"os-R\'enyi (ER) networks and on lattices with 
dimensions 2, 3, 4, and 5. We obtain new and surprising results in all these cases, and we correct statements 
in the literature for ER networks and for 2-d lattices. In particular, we find that $d=4$ is the upper critical
dimension, that the percolation transition is continuous for $d\leq 4$ but -- at least for $d\neq 3$ -- not 
in the universality class of ordinary percolation. For ER networks  we verify that the cluster statistics is 
exactly described by mean field 
theory, but find evidence that the cascade process is not. For $d=5$ we find a first order transition as for 
ER networks, but we find also that small clusters have a nontrivial mass distribution that scales at the 
transition point. Finally, for $d=2$ with intermediate range dependency links we propose a scenario different
from that proposed in W. Li {\it et al.}, PRL {\bf 108}, 228702 (2012).
\end{abstract}

\maketitle

\section{Introduction}

While percolation and related epidemic processes had appeared until recently as a mature subject that holds 
few surprises, this has changed dramatically during the last few years \cite{Araujo}. A host of new models have been 
proposed, such as percolation on growing networks \cite{Callaway}, percolation on hierarchical structures
\cite{Singh}, agglomerative percolation \cite{Bizhani,Lau}, explosive percolation \cite{Achlioptas}, 
percolation on interdependent networks \cite{Buldyrev}, percolation where nodes cooperate to infect their
neighbors \cite{Dodds,Janssen,Bizhani2}, and percolation where spreading agents cooperate \cite{Chen,Cai}.
Indeed, some of these models are not entirely new and are related to models studied since the 1970's.
The model of \cite{Singh}, e.g., can be viewed as a version of percolation on lattices with long range 
contacts \cite{Aizenman,Grass2013}, while cooperative percolation \cite{Dodds,Janssen,Bizhani2} can be
viewed as a variant of bootstrap percolation \cite{Chalupa,Adler,Goltsev} \cite{Note1}
 with heterogeneous nodes \cite{Baxter}.
But these models had previously been widely considered as curiosities, while only the recent developments
have shown their wide range and wide spread applicability.

The range of behaviors found in these models is bewildering. Instead of the continuous
(``second order") transition with standard finite size scaling (FSS) observed in ordinary percolation (OP),
one finds everything from infinite order transitions with Kosterlitz-Thouless (KT) type scaling \cite{Callaway}
to first order transitions with KT type scaling \cite{Grass2013} to first order hybrid transitions 
\cite{Goltsev,Baxter,Bizhani2,Cai}, and -- last but not least -- second order transitions 
with completely different FSS behavior \cite{Grass2011}.

Some of these results were obtained by simulations, others were derived analytically.
With very few exceptions \cite{Callaway,Singh}, the latter rely on the fact that mean field theory becomes exact 
on random locally tree-like networks \cite{NSW,Newman,Bollobas}. The latter allows very elegant treatments
based on self-consistency equations derived using message passing arguments,
as e.g. demonstrated in \cite{Goltsev,Baxter,Bizhani} for cooperative percolation and in \cite{Son2} 
for several interdependent network models \cite{Buldyrev,Parshani,Bashan,Gao}. But these results can be very 
deceptive -- as seen by simulations \cite{Son1,Cai} -- when applied uncritically to networks which are either 
not random or not (locally) tree-like.
For such networks dynamic message passing \cite{Karrer-a,Karrer-b,Shrestha} and the cavity 
method \cite{Shiraki,Watanabe} have been applied recently with very promising 
results, but their applicability to most of the above models is still far from 
obvious. The fact that network topology can change the critical behavior entirely was demonstrated 
recently \cite{Cai} for cooperative spreading agents (so-called syndemics or co-infections).
Another warning should be the fact that interdependent two- and three-dimensional lattices do not 
show the first order transition found on random networks, but show second order transitions \cite{Son1},
if all links are short range.

All this shows that efficient methods to simulate such models are badly needed. While there exist 
very efficient methods for OP and for all versions of cooperative percolation, this is not the case 
for the class of models introduced in \cite{Buldyrev} and developed further in 
\cite{Parshani,Bashan1,Bashan,Gao,Li-Bashan,Berezin1,Berezin2,Baxter-b,Watanabe,Son1,Son2,Valdez,Cellai,Stippinger,Danziger,Zhou}.
For these models which can either be viewed as describing cascading failures on {\it interdependent networks}
or {\it viable clusters on multiplex networks},
the algorithms used so far in large simulations are extremely slow.
After the present work was done, we learned of a recent algorithm \cite{Hwang} that is fast but highly 
non-trivial, and which has so far not yet been applied to any real problem \cite{Note2}.
It is the purpose of the 
present paper to present such an algorithm and to use it for several different network topologies. 
As we shall see, the results are most disturbing, as the behavior is different for each case. 
This shows that one should be extremely cautious in applying results obtained mathematically for 
random locally tree-like models to real world situations.

In deriving the algorithm we use a mapping of the problem onto a solid-on-solid (SOS) \cite{Privman} type 
growth model. This mapping is also of interest by itself, as it shows that the model has a number of 
non-trivial features that might become useful in future analytic treatments.

In the next section we shall define the model more formally, discuss interpretational differences with
\cite{Buldyrev} and the mapping onto an SOS type model, and present our fast algorithm. Applications to 
Erd\"os-R\'enyi networks are treated in Sec.~3A, where we shall present arguments that the static
properties of viable clusters are indeed as described by mean field theory, but not the dynamics of 
cascades. Applications to regular lattices in 2 to 5 dimensions are presented in Secs.~3B to 3E. 
In particular we shall show that the percolation transition for $d=2$ is not in the universality class
of ordinary percolation, and that $d=4$ is an upper critical dimension. Finally, in Sec.~3F we shall
discuss the behavior on 2-dimensional lattices with intermediate range links. The results are summarized,
and open problems are pointed out, in the final Sec.~4.

\section{The model, its mapping onto SOS type growth, and the resulting algorithm}

The models studied in 
\cite{Buldyrev,Parshani,Bashan1,Bashan,Gao,Li-Bashan,Berezin1,Berezin2,Baxter-b,Watanabe,Son1,Son2,Valdez,Cellai,Stippinger,Danziger,Zhou} 
were originally presented as interdependent networks showing failure cascades subsequent to random removal 
of nodes, but as noted in \cite{Son1,Son2} they are more easily interpreted as single multiplex networks. More 
precisely, while in general interdependent networks some nodes depend on each other, these dependencies 
always were assumed in most of these papers to be pairwise and mutual. In this case each pair of interdependent 
nodes can be identified into a single node, and the network becomes just a duplex network, i.e. one set of nodes 
connected by two sets of (undirected) links. In some other papers \cite{Gao,Bashan,Son2} this was generalized 
to $m$ networks with mutual dependencies among all nodes in clusters of sizes $\leq m$, where each cluster contains
at most one node per network. In this case identification of all nodes in such 
clusters leads to $m$-plex networks. In the following we shall only consider the case $m=2$, i.e. duplex
networks or pairwise mutual dependencies.

Another slight but important shift 
was made by \cite{Son1,Son2} when they noted that the model can be formulated as a self-consistency
problem without any reference to node removals (``damage") and to cascades triggered by them. 
Rather, as also pointed out in \cite{Min-Goh},
it describes (for $m=2$) the case where each node needs two essential supplies for being active.
More precisely, we assume that there exist a source node that supplies both resources, and that the 
resources can be transported only on active nodes. It is this latter interpretation which we adopt: 
We usually consider a cluster $\cal C$ of nodes as {\it viable}, if any two of its nodes are connected by paths 
on both sets of links -- where we also demand that both paths are entirely confined to $\cal C$. In that case,
any node in $\cal C$ can be supplied with both resources, if the source node is also in $\cal C$. Conversely,
if any node in $\cal C$ can obtain both resources, i.e. if any node in $\cal C$ is doubly connected to the 
source, then (by the assumed undirectedness of the links) also two arbitrary nodes in $\cal C$ are 
doubly connected. Notice that in this way we do not consider the networks themselves as 
failed or intact, but we consider only the activities {\it on the networks} as present or absent. To use 
the main example used in \cite{Buldyrev}: If there is a power failure, the electricity network itself might 
still be perfectly intact, it is just the activities on the power grid and on the associated computer 
network that have gone down.

The next difference with the bulk literature on interdependent networks is that we do not in general delete 
nodes, but we consider the fate of viable clusters as we change the densities of links. On ER lattices,
decimating nodes is just equivalent to renormalizing the connectivity and thus equivalent to decimating links 
\cite{Son1}. On regular lattices this is not true, but keeping all nodes just reduces the number of control
parameters by one. For two-dimensional (2-d) lattices we checked, however, that decimating both sites and 
links leads to the same universality class.

A last difference with \cite{Buldyrev} is that we consider not only the largest viable cluster
on the entire network, but we study {\it all} viable clusters, in agreement with \cite{Hwang}. As we shall 
see, this comes with no additional effort. Indeed, only by studying all viable clusters we can verify with
our algorithm which one is the largest one. Again we consider this as more realistic. To stay with the 
electric breakdown example: Assume we have a power failure in central Italy which disconnects the north 
from the south. This would be considered by \cite{Buldyrev} as catastrophic, since there would no 
longer exist a giant viable cluster. But as long as there local viable clusters around Milano and Napoli, 
people there would be perfectly happy.

In spite of all the differences with \cite{Buldyrev} mentioned in this section, notice that we are 
mathematically still dealing with precisely the same class of models, and expect the same type(s) of phase 
transitions.

Our algorithm works for any type of duplex network with undirected links, although the mapping onto an 
SOS growth model requires
of course strictly spoken that we deal with a planar lattice (since SOS models are models for 2-d surfaces).
We shall use therefore a language adequate for 
this special case, although it should be understood that everything in the following applies also the general
case. We thus have a set of $N$ points which are partially connected by two sets of bonds (``red" and ``green", say)
between neighboring sites. Bonds of either color are placed randomly and independently, with probability 
$q \in [0,1]$.

Our algorithm has two ingredients. In the first part we pick a seed (or ``source") site and find the largest 
viable cluster connected to this seed. In the second part we repeat this for all possible seeds.

{\bf (a) Finding the largest viable cluster ${\cal C}$ attached to point $i$:} This is simply done by 
alternatively performing ``epidemic" or ``Leath-type" \cite{Leath} spreading processes on the red and 
green bonds, each starting from site $i$ (this can be done 
breadth or depth first; we actually used breadth first. We also assume for definiteness that we start 
with the red bonds). We do not fix bond occupancies "on the run", but 
we rather determine them before we start with the first epidemic. We follow
the spreading until it dies due to the finiteness of the lattice, which gives us a first cluster ${\cal C}_1$.
Since ${\cal C}_1$ is connected to the seed site but not necessarily doubly connected, we know that 
the largest viable cluster attached to site $i$ must be contained in it, ${\cal C}\subseteq {\cal C}_1 $. 
Therefore, when we generate the second epidemic using the green bonds, we restrict ourselves to ${\cal C}_1$, 
generating thereby a cluster ${\cal C}_2$ with ${\cal C}\subseteq {\cal C}_2 \subseteq {\cal C}_1 $. As we proceed 
alternatingly, we generate thus a chain of nested sets
\be 
  {\cal C} \ldots \subseteq {\cal C}_{h} \subseteq {\cal C}_{h-1} \ldots \subseteq {\cal C}_1 .
\ee

\begin{figure}[htp]
\includegraphics[scale=0.30]{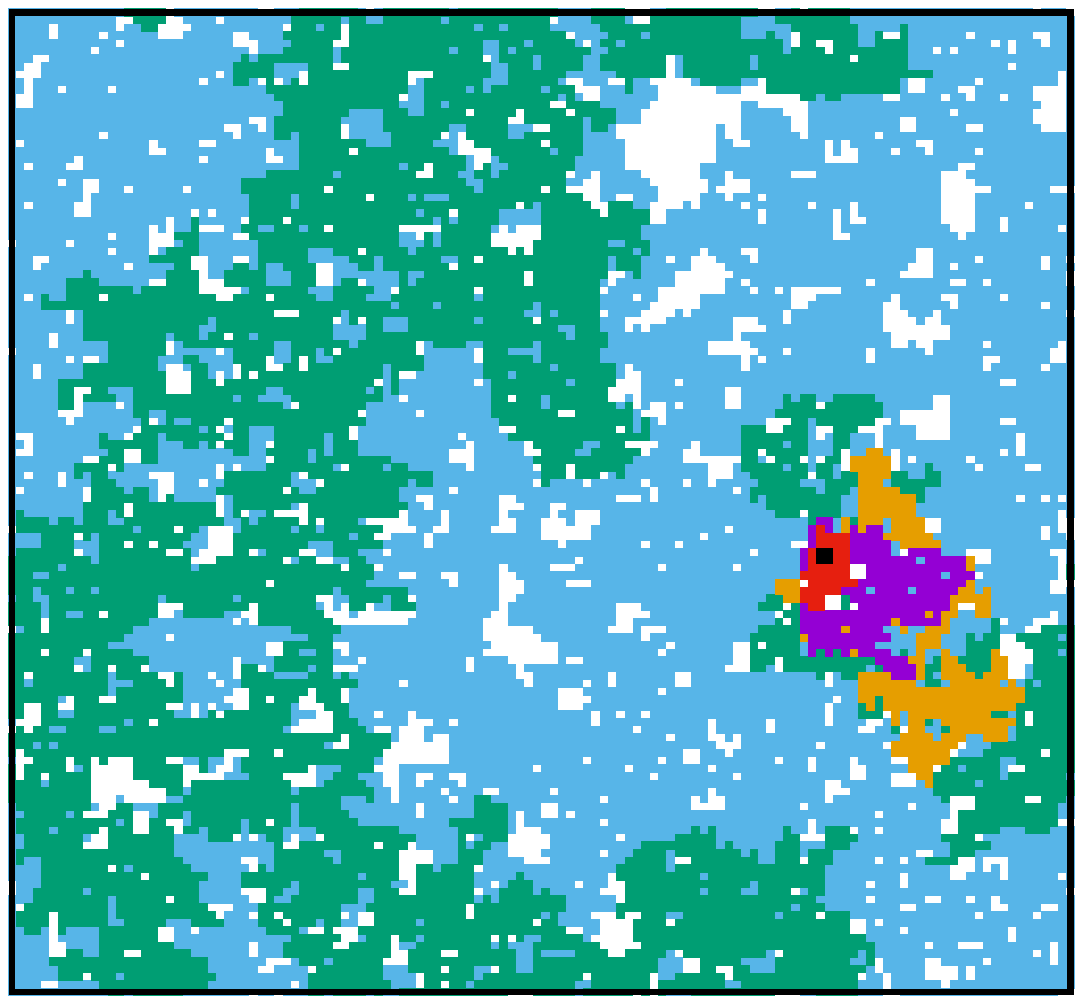}
\includegraphics[scale=0.30]{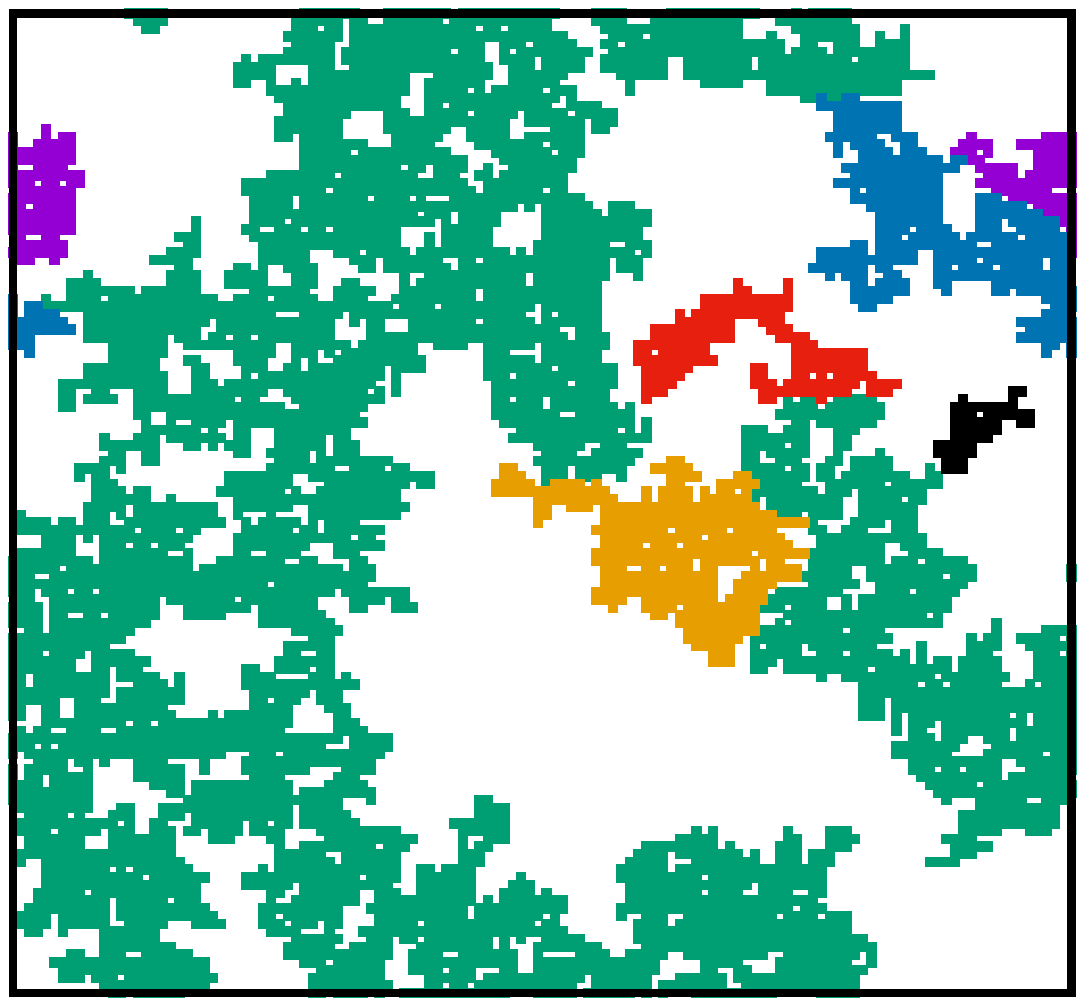}
\caption{(color online) panel (a): Clusters ${\cal C}_1$ (all non-white points) to ${\cal C}_5$ (dark blue) 
   attached to the site $i$ indicated by the black dot. Each color corresponds to one terrace, with light blue $<$
   green $<$ orange $<$ magenta $<$ red indicating increasing heights; panel (b): six large clusters of 
   height 2, all attached to points in the cluster ${\cal C}_1$ shown in panel (a). Notice that all these clusters 
   are fully contained in ${\cal C}_1$, and no two clusters overlap partially. They are either disjoint, or one 
   is a subset of the other. Notice also that some of these clusters touch each 
   other, indicating the possibility that some adjacent terraces may have equal heights. The control parameter 
   is set such that ordinary percolation would be supercritical, but mutually interdependent percolation of 
   viable clusters is subcritical.}
\label{fig.0}
\end{figure}

Since the lattice is finite, this must stop at a finite value of $h$ which we call $h_i$. It is easy to see
that $ {\cal C}_{h_i}$ is equal to ${\cal C}$. If not, then ${\cal C}_{h_i}$ would contain at least one 
point which is not connected to the seed by paths of either color, but in that case ${\cal C}_{h_i+1}$ would
be strictly larger than ${\cal C}_{h_i}$, which is in conflict to our assumption that the iteration stops
at $h_i$. 

In the following, we shall call each epidemic starting from $i$ a ``wave", and $h$ its {\it height}. Since 
the supports ${\cal C}_h$ of successive waves are nested, we obtain in this way a landscape with a single 
mountain of peak height $h_i$, and ``terraces" of heights $h < h_i$ (see Fig.~1). Each terrace is just 
\be
    \Delta {\cal C}_h \equiv {\cal C}_h \setminus {\cal C}_{h+1}.
\ee
 The part of the lattice which was 
not even touched by the first wave is called the zeroth terrace with height $h=0$ \cite{Note5}.

{\bf (b) Finding viable clusters attached to other points:} After we have obtained the largest cluster attached 
to $i$ we could pick another site $j$ (either randomly or by going systematically through the lattice), and 
repeat the same process. To find the largest viable cluster on the entire lattice ${\cal C}_{\rm max}$, we would 
then have to repeat this for all sites. But this would be extremely time consuming. We are interested in the phase 
transition where ${\cal C}_{\rm max}$ becomes macroscopic in the limit $N\to\infty$. This happens at a 
critical value $q_c$ of $q$ which is far above the critical value for single epidemics. 
Thus for all seeds the first few waves will most likely be huge, and most of 
the CPU time is spent for finding out clusters which overlap so much that they are nearly identical and 
 cover nearly the entire lattice. The resulting CPU time would then be roughly ${\cal O}(N^2)$.

Fortunately, there is no need to use this brute force algorithm. Let us consider the system of waves and
terraces for the second seed $j$. Since the entire lattice is covered by terraces from the first seed $i$,
we can assume that $j$ is on a terrace of some height $h \in \{0,1,\ldots h_i\}$. If the height were equal to $h_i$, 
$j$ would be part of the first cluster. Thus we can assume that $h < h_i$. If $h=0$, then $j$ is not connected to $i$
by the red bonds, and thus any viable cluster attached to $j$ must be contained in the zeroth terrace 
$\Delta {\cal C}_0$. Otherwise (if $0<h<h_i$) we can use

{\bf Lemma 1:} {\it If $h>0$, the first $h$ waves starting from $j$ cover exactly the same clusters ${\cal C}_{h'}$ (with $h'\leq h$)
as the waves starting from $i$. }

{\bf Proof:} 
The proof is by induction and uses the fact that each wave just covers all sites on a given network that can be reached 
from the seed, where each of these networks is the subgraph of the original network reached by the previous wave. 
First of all, it is clear that the lemma holds for $h=1$, because ${\cal C}_1$ is the set of all points on the original
network that are connected to the seed by red bonds. Since the two seeds are connected by red bonds due to the assumption that $h>0$, 
any point $k$ connected to $i$ must also be connected to $j$ and vice versa.
Let us now assume that for some $h'<h$ the clusters attached to $i$ and $j$ are the same. Then we can use exactly the 
same argument, just replacing the original network by ${\cal C}_{h'}$ and using the appropriate color of the bonds.

Thus we do not need to follow the first $h$ waves starting from $j$, and we can immediately start with wave $h+1$.
In doing so, we can use also another important simplification because of 

{\bf Lemma 2:} {\it Assume that point $j$ is on a terrace of height $h < h_i$. Then the entire next wave is confined to
this terrace.}

{\bf Proof:}  
For the proof we have to show that the next wave neither spills over to the lower terrace, nor to the higher.\\
For the first part we can assume $h>0$. Assume now that the wave actually does spill over, i.e. there exists a 
point $k$ which is connected to $j$ but is on a lower terrace. This point would not be connected to $i$ on any terrace
with $h'>=h$, while $j$ is. This cannot be.\\
For the second part we assume for the moment that the wave spills over to a higher terrace, i.e. there exists a point 
$k$ which is connected to both $k$ and $i$. But since $j$ is not on this higher terrace (i.e. is not connected to $i$), 
this cannot be either.
 
Thus when building the terraces for site $j$ we can restrict ourselves to waves for which all boundaries of the 
terraces generated by the first seed form obstacles which cannot be crossed. It is easy to see that this generalizes
also to all later seeds, when the landscape is made up be any number of terraces and boundaries between them:

{\bf Proposition 1:} {\it If a seed $j$ happens to be on any terrace with a height $h$ which is not a local maximum
(i.e., $j$ is not in a locally maximal viable cluster), then all previous wave boundaries form obstacles for 
all avalanches starting at $j$ which have to be followed in order to get the viable cluster attached to $j$.}

The proof uses the same arguments as above. We just have to realize that successive waves that started from later and 
later seeds are just distinguished by more and more restricted subsets
of the original network to which they are confined. Whatever these subsets are, the above proofs go through without
modification.

It should now be clear why we call this procedure an SOS type growth process. As in any SOS model, we have a 
rough landscape and the growth of this landscape proceeds by localized events, each of which builds a hierarchy 
of nested terraces. As in surface diffusion problems with strong Schwoebel barriers, these events cannot spill 
over the boundaries set by previous events. 

Notice that neighboring terraces belonging to different seeds can have 
the same height. Also in this case, their boundaries cannot be crossed by later waves from other seeds. To implement 
this restriction we either use an additional marker for each site which tells us the seed to which it belongs; or,
alternatively, when each wave is finished, we cut all bonds connecting the wave with its complement. Both methods
allow us to prevent the waves from spilling over the terrace boundary, without modifying the dynamics inside any terrace.
Also, when we start with a new
seed $j$, we choose the color of the first wave according to the landscape height $h$ at this seed, and we count subsequent 
heights by adding to this $h$. Since now every site $i$ is ``infected" precisely by $h_i$ waves, it follows that
the algorithm has time complexity $N \langle h\rangle$, where $\langle h\rangle$ is the average height.
This estimate holds for arbitrary graphs, provided all node degrees are bounded.

Before we show numerical results, we present also a second proposition which we did not find useful for numerics,
but which could be very useful for mathematical treatments.

{\bf Proposition 2:} {\it The final landscape and the system of all terrace boundaries is independent of the 
sequence by which the seed points are chosen. }

Thus the landscape is not a property of the realization of the algorithm (which involves an arbitrary choice 
of going through all seeds), but is an inherent property of the network. In essence, it says that the growth 
of the surface is Abelian in a similar way as the 
Bak-Tang-Wiesenfeld sand pile model is Abelian \cite{Dhar}.

For the proof we can use the fact that any permutation of sites can be written as a product of 
pairwise transpositions. We thus need to show only that it does not matter whether we first take seed $i$
and then seed $j$ or the inverse. But this follows from the arguments in the proofs of Lemma 1 and 2. 

We should add that we checked both propositions also numerically, finding perfect agreement.

Finally, we should point out how our sequences of waves are related to the failure cascades of \cite{Buldyrev}
and to other algorithms proposed for them.
One essential difference is that we do not demand that the cluster surviving the cascade is always the maximal
one. Rather, the maximal cluster is determined at the end, when all viable clusters are known. This simplifies 
the algorithm of course considerably. Also the complexity of the algorithm of \cite{Hwang} is related to 
the fact that they always follow the largest viable cluster. This precaution is not taken in the algorithm 
of \cite{Schneider}, where it is {\it assumed} that a cluster which starts large at the beginning of the cascade 
is not overtaken later by one which starts smaller. This is true for sparse tree-like graphs (whence their
algorithm gives correct results), but it is not true in general. Related to this is the 
absence of the notion of `seed' or `source' in these algorithms. While in our algorithm each `cascade'
gives just one cluster arbitrarily picked by the seed, the cascades studied in \cite{Buldyrev} are constructed so as
to lead always to the largest cluster. As a consequence, the average cascade length in \cite{Buldyrev} is
in our interpretation essentially the average height of the largest viable cluster, which is in general
(but not always?) an upper bound for the average surface height.

A last difference between our algorithm and that of \cite{Buldyrev} is that we always start our `cascades'
from the full (undamaged) system, while the cascades of \cite{Buldyrev} are supposed to be triggered by 
(small and successive?) damages, starting from an already partially damaged system. On the other hand, our 
algorithm shares with that of \cite{Buldyrev} that it does {\it not} deal -- in contrast to what is suggested
in \cite{Buldyrev} -- with any real dynamics. Both algorithms deal with pseudodynamics in a fictitious time, 
and should not be confused with actual cascading processes going on in real time.

\vspace{.6cm}

\section{Applications}

\subsection{Erd\"os-R\'enyi Networks}

We first apply our algorithm to duplex ER networks where both layers have the same average degree 
$z = \langle k \rangle$. In this case it is known from \cite{Son1,Son2} that the generating function formalism 
of \cite{NSW,Newman} gives the exact threshold and the exact 
dependency of the order parameter on $\langle k \rangle$. Thus it can be used to test the accuracy 
of the algorithm. On the other hand, we shall also use it to find the average height of the landscape and 
the average height of its maxima (i.e., of the largest clusters). The latter is just the average 
cascade life time in the interpretation of \cite{Buldyrev}. Since during the cascades the clusters
are supercritical, it is not a priori clear whether the mean field theory arguments used in \cite{Buldyrev}
to derive a scaling law for this life time are exact. To make a short summary, we will find that the 
algorithm works perfectly, but the mean field arguments of \cite{Buldyrev} for the life time seem to be 
to be only approximate. This implies that also the arguments of \cite{Buldyrev} in favor of the 
percolation threshold value have to be taken with care -- although they give the correct result --
because they rely on the assumed mean field cascade dynamics.

\begin{figure}[htp]
\includegraphics[scale=0.31]{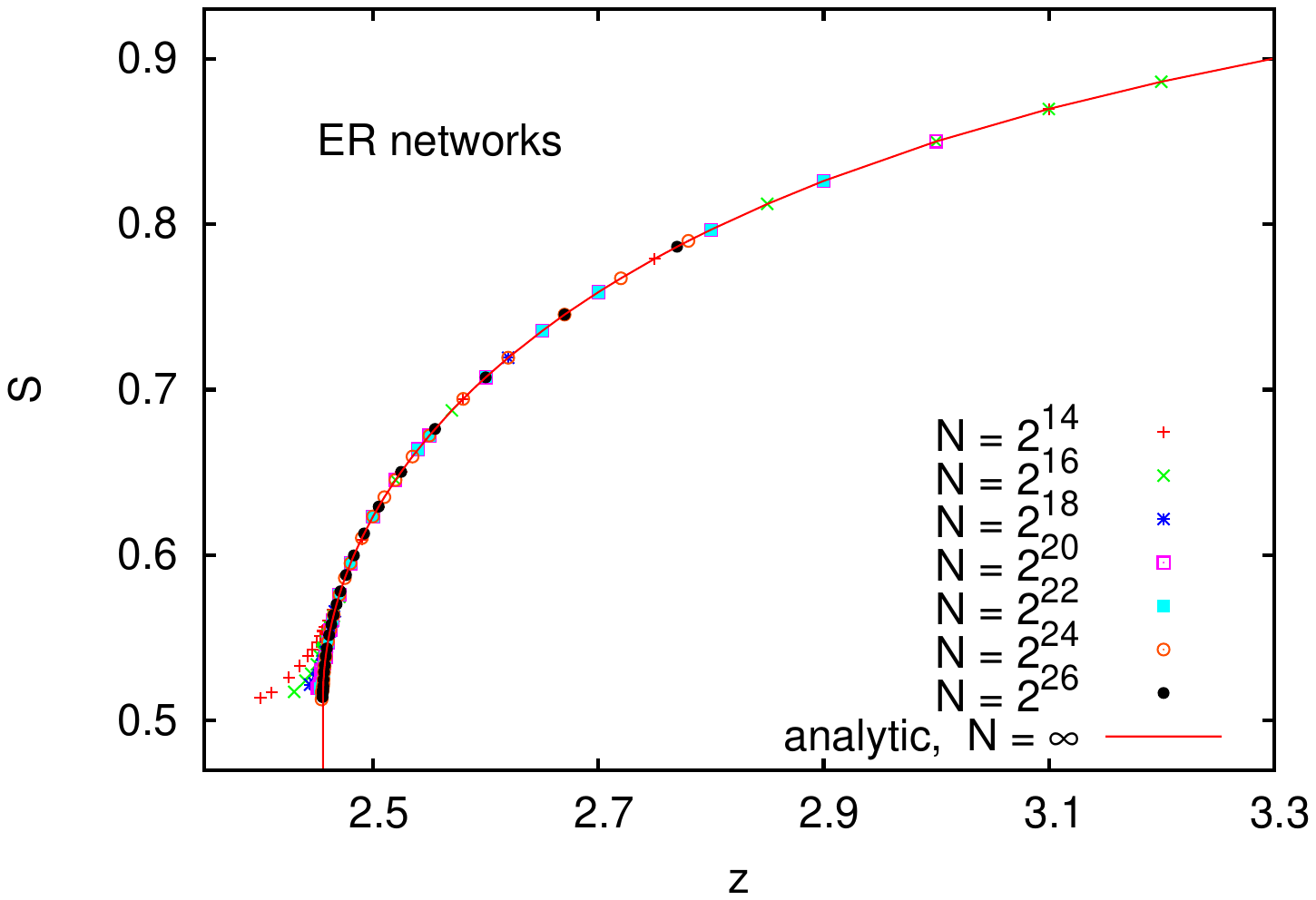}
\caption{(color online) Order parameter (density of giant viable clusters) for ER networks plotted against the
   average degree $z$. The continuous curve is the prediction of Eq.~(\ref{S-ER}), while the points are from 
simulations with $N$ ranging from $2^{14}$ to $2^{26}$. For $S<0.511700...$ the theoretical curve is strictly 
vertical, while it has square-root behavior above. }
\label{fig.1}
\end{figure}

\begin{figure}[htp]
\includegraphics[scale=0.31]{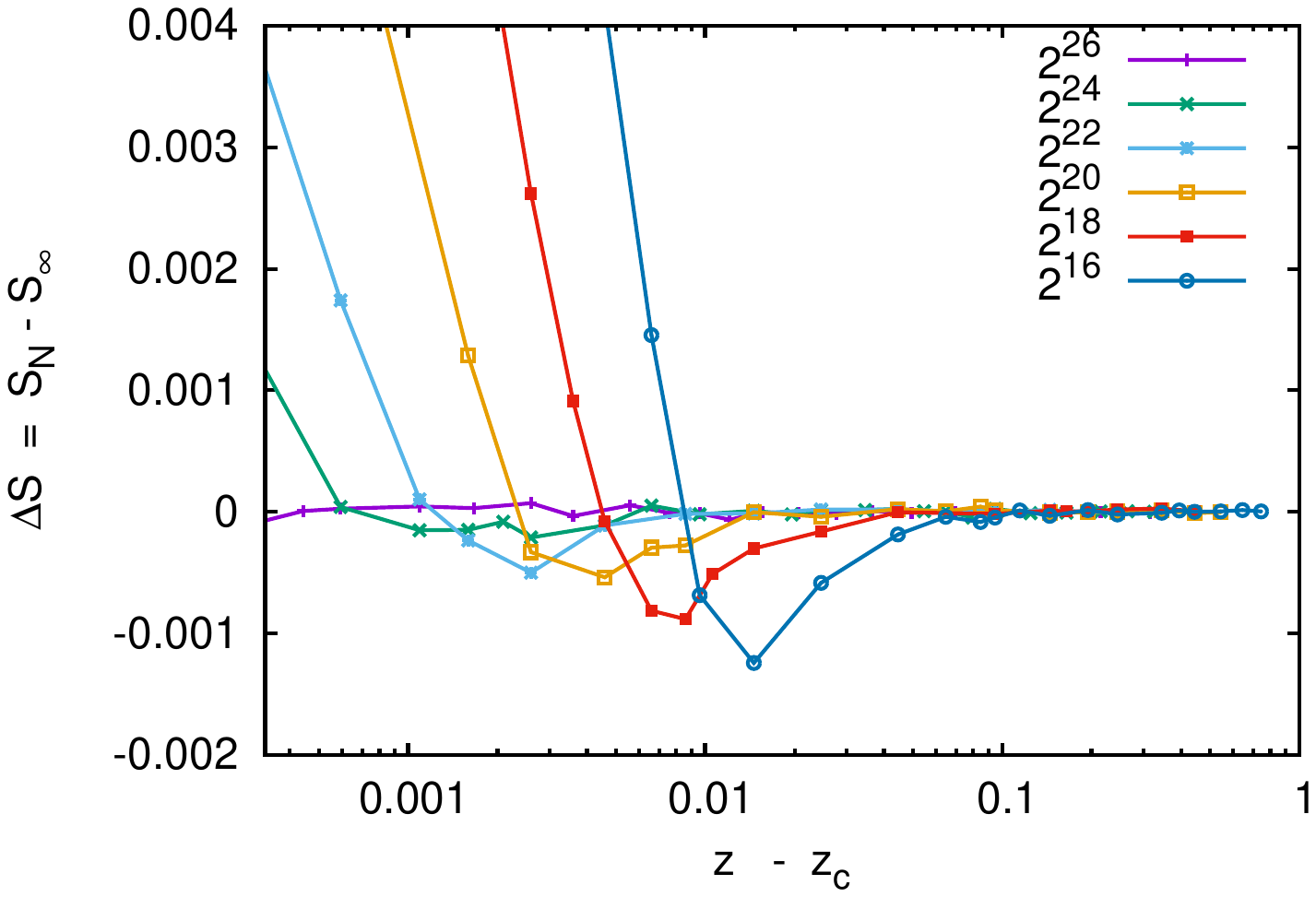}
\caption{(color online) Differences between the data points and the theoretical curve in Fig.~2. To enhance the 
region $z\approx z_c$, the data are plotted against $z-z_c$ on a logarithmic axis.}
\label{fig.2}
\end{figure}

\begin{figure}[htp]
\includegraphics[scale=0.31]{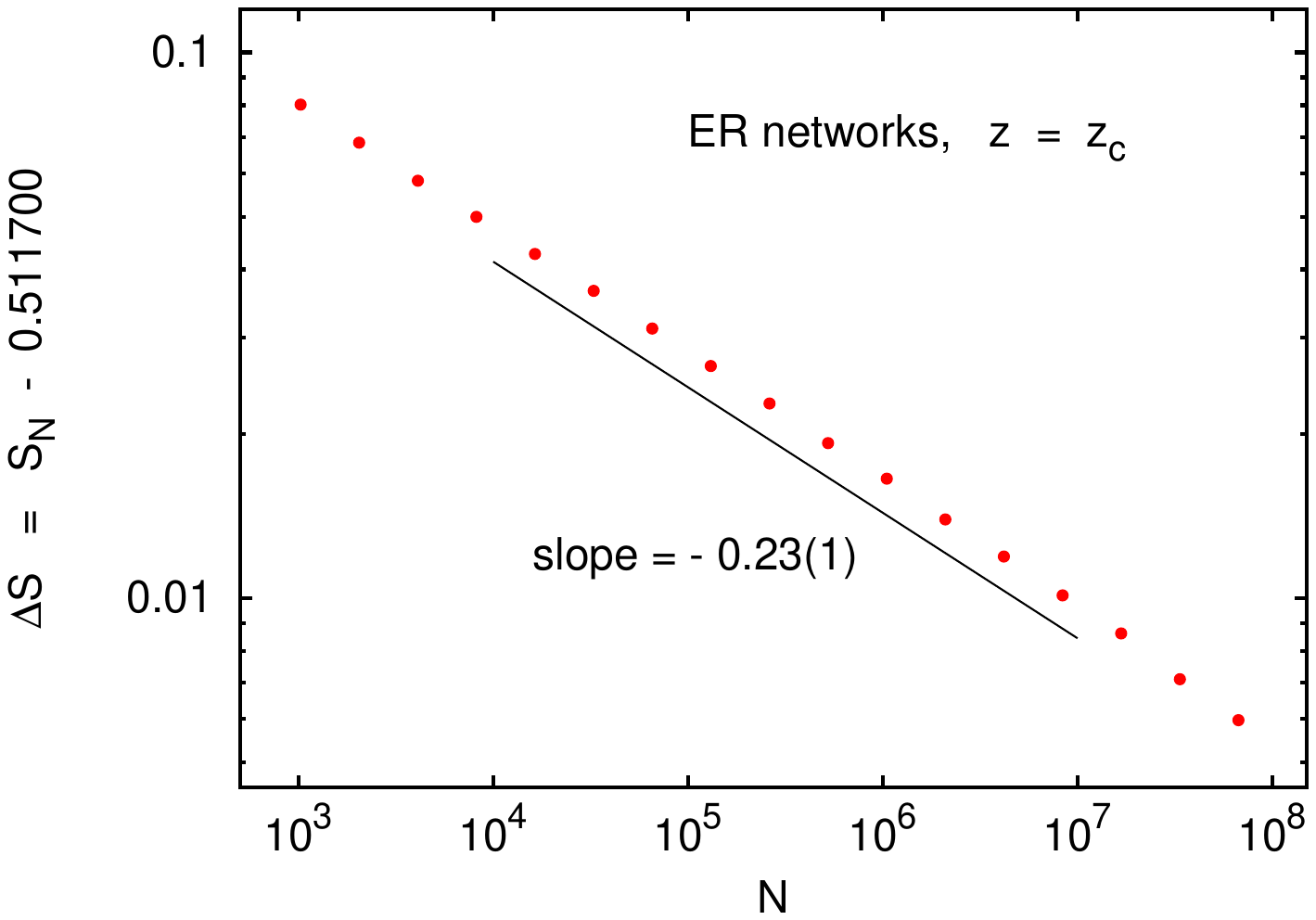}
\caption{(color online) Differences between the data points for $z=z_c$ and the theoretical value 0.511700 towards
which they should converge for $N\to\infty$. We see indeed a straight line on a log-log plot, which leads to 
Eq.~(\ref{ER-S_z_c}).}
\label{fig.3}
\end{figure}

We consider networks with $N=2^m$ nodes, with $m = 4,\ldots 26$. The average degree was varied in the range 1.9 to 
3.2. The critical point is known in this case to be at $z_c = 2.45540748\ldots$. The 
number of realizations simulated at $z_c$ varied from $\approx 10^8$ for $N=1024$ to $>600$ for $N=2^{26}$.

The density of the giant viable cluster on an infinite ER network is given by the largest solution of the equation \cite{Son1}
\be
    F(S) \equiv  S -(1-e^{-zS})^2 = 0.                        \label{S-ER}
\ee
It vanishes for $z<z_c$, and has both a jump and a square root singularity at $z=z_c$,
\be
    S = 0.511700 + a \sqrt{z-z_c} + ... \quad {\rm for} \quad z>z_c.
\ee
For finite $N$ there are of course corrections, the analytic form of which is not known.

Numerical results for $S_N$, the average relative size of the giant cluster on a graph of size $N$, are given 
in Figs.~2 to 4. Actually, in these plots we show (in contrast to analogous plots in the next subsections) $S_N$ 
conditioned on realizations which do have a giant cluster. 
Notice that all other clusters are extremely small (for $N>2^{20}$ we found only clusters of sizes
1 or 2 apart from the giant one, and even for $N=1024$ the largest non-giant clusters had sizes $<8$), thus it 
is straightforward to identify giant clusters, except for $N<10^3$. In Fig.~2 we show $S_N$ versus $z$. Since 
this gives a perfect agreement with Eq.~(\ref{S-ER}) except for $z$ extremely close to $z_c$, we plot in Figs.~3 and 4
differences between analytical and numerical results for a more sensitive comparison. From Fig.~3 we see that the
finite-$N$ corrections are indeed not monotonic, as one might have suggested from Fig.~2, but they decrease fast 
with $N$. For $N=2^{26}$, the deviations from the theory for $N=\infty$ are $<10^{-4}$ for all $z$ except extremely close 
to $z_4$, which is just the level of statistical errors. Thus we verified with very high precision that our algorithm 
gives results in agreement with Eq.~(\ref{S-ER}) not only near the critical point, but also for all $z>z_c$. The agreement
at $z=z_c$ is checked in Fig.~4, where the differences $\Delta S_N = S_N - S_\infty$ are plotted against $N$. We see 
a perfect scaling law 
\be
   \Delta S_N \sim N^{-\alpha} \quad {\rm with} \quad  \alpha = 0.23(1).    \label{ER-S_z_c}
\ee
To our knowledge there is no theoretical prediction for $\alpha$. For $z>z_c$, Fig.~3 is consistent with
$\Delta S_N \sim N^{-\alpha} \phi[(z-z_c)N^{1/2}]$, but the data are too noisy make a strong claim for it.

\begin{figure}[htp]
\includegraphics[scale=0.31]{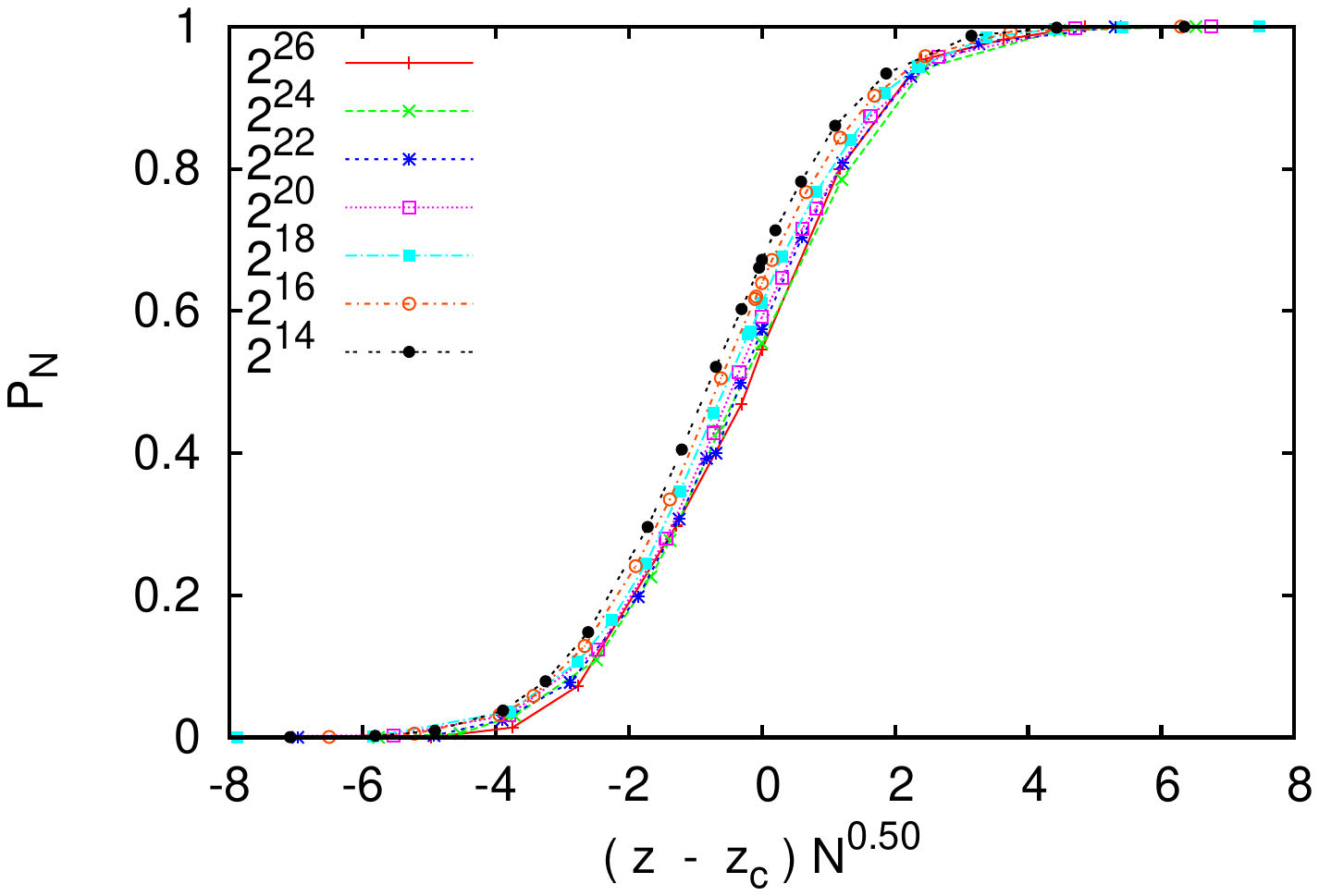}
\caption{(color online) $P_N(z)$, the probability to have a giant viable cluster on an ER network of size $N$, plotted
against $(z-z_c) N^{1/2}$.}
\label{fig.4}
\end{figure}

There also seems to exist no prediction for the behavior of $P_N(z)$, the chance that there exists a viable 
giant cluster on a network of size $N$. In analogy to ordinary percolation we expect that this is a step function in the 
limit $N=\infty$, i.e. $P_\infty(z) = \Theta(z-z_c)$, but the behavior for finite $N$ is non-trivial. In Fig.~5 
we plot $P_N(z)$ against $(z-z_c) N^{1/2}$. We see that all curves become essentially parallel, i.e. we have a 
data collapse apart from a correction to scaling that leads to a shift of the effective transition point towards higher 
$z$ as $N$ increases.

\begin{figure}[htp]
\includegraphics[scale=0.31]{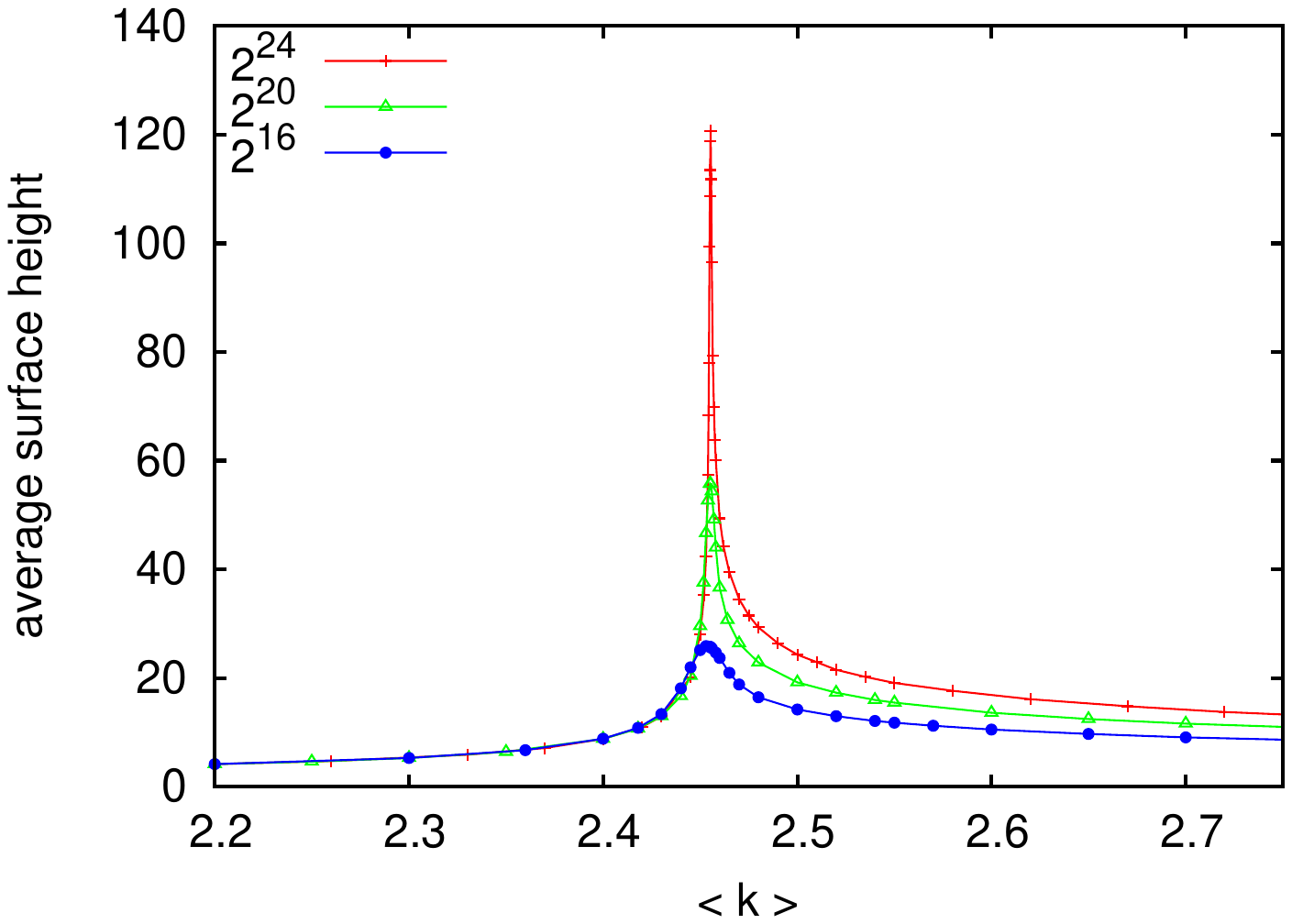}
\caption{(color online) Average ``surface" heights (i.e. cascade life times) for three network sizes, plotted against 
$z$. While the collapse of the three curves for $z \ll z_c$ is easy to understand, the behavior at larger $z$ is more 
interesting.}
\label{fig.5}
\end{figure}

``Surface" heights (i.e. average cascade life times) measured in these simulations are discussed in the next figures.
In Fig.~6 we show for three different network sizes how the average surface heights, averaged over all nodes, depend
on $z$. As expected, they are small for $z \ll z_c$ and become quickly independent of $N$: For small $z$ there are 
only small clusters to start with, and the cascades terminate quickly. For $z \gg z_c$ there are only small holes 
in the viable clusters and they also are quickly identified, whence the cascades terminate soon also in this regime.
It is less obvious why the heights do not seem to become independent of $N$ as $z$ becomes very large. The most 
conspicuous feature of Fig.~6 is, however, the sharp peaks developing at $z=z_c$ as $N$ increases. Figures 7 to 9
deal with several aspects of these peaks.

Heights of the peaks (more precisely the average surfaces heights at $z=z_c$; the peaks occur slightly to the left of $z_c$) 
are shown in Fig.~7. In the same figure we also show, in addition to the {\it average} surface heights, the 
(averages of the) surface peaks, i.e. the highest points in the ``landscapes". At least for ER networks (where we 
have no large viable clusters apart from the giant ones) we expect these peak heights to be precisely the 
average durations of the cascades studied in \cite{Buldyrev} (see also \cite{Zhou}). We see from Fig.~7 that 
the {\it average} heights show a perfect scaling,
\be
   \langle h\rangle \sim N^{\gamma}\;,\qquad \gamma = 0.280(1).    \label{ER-h_N}                           
\ee
In contrast, the peak heights in the landscape seem to increase with the same asymptotic 
power law, but with large finite-$N$ corrections. The latter are presumably the usual bias corrections in extremal
properties evaluated over finite domains. One has similar (logarithmic?) corrections e.g. in the average size 
of the largest cluster in subcritical (ordinary) percolation on finite lattices.

Eq.~(\ref{ER-h_N}) is in clear contradiction to the result $h_{\rm peak} \sim N^{1/4}$ obtained in \cite{Buldyrev}
by mean field arguments which do give also the correct Eq.~(\ref{S-ER}) for the cluster size and the exact value 
of $z_c$. This is rather puzzling. One might try to explain it by noting that the value of $z_c$ depends only 
on clusters which are barely viable, and thus the local tree-likeness of sparse ER networks should be sufficient --
while the cascade dynamics deals always with supercritical clusters. But then it is not clear why Eq.~(\ref{S-ER})
should be correct also for $z>z_c$, which we took great pain to verify. Notice that the violation of the mean field
prediction seen in Fig.~6 is not related to the observations in \cite{Zhou}. In that paper it was shown that 
an apparent violation of mean field theory is observed, if one does not use the true critical value of $p_c$
in the analysis, but an effective one that changes from realization to realization. In our analysis, $p_c$ is 
always the true critical point. In any case, our value of $\gamma$ disagrees both with
the mean field exponent 1/4 obtained in \cite{Buldyrev} and with the exponent 1/3 proposed in \cite{Zhou}, 
which was based on simulations of much smaller systems with much lower statistics.

\begin{figure}[htp]
\includegraphics[scale=0.31]{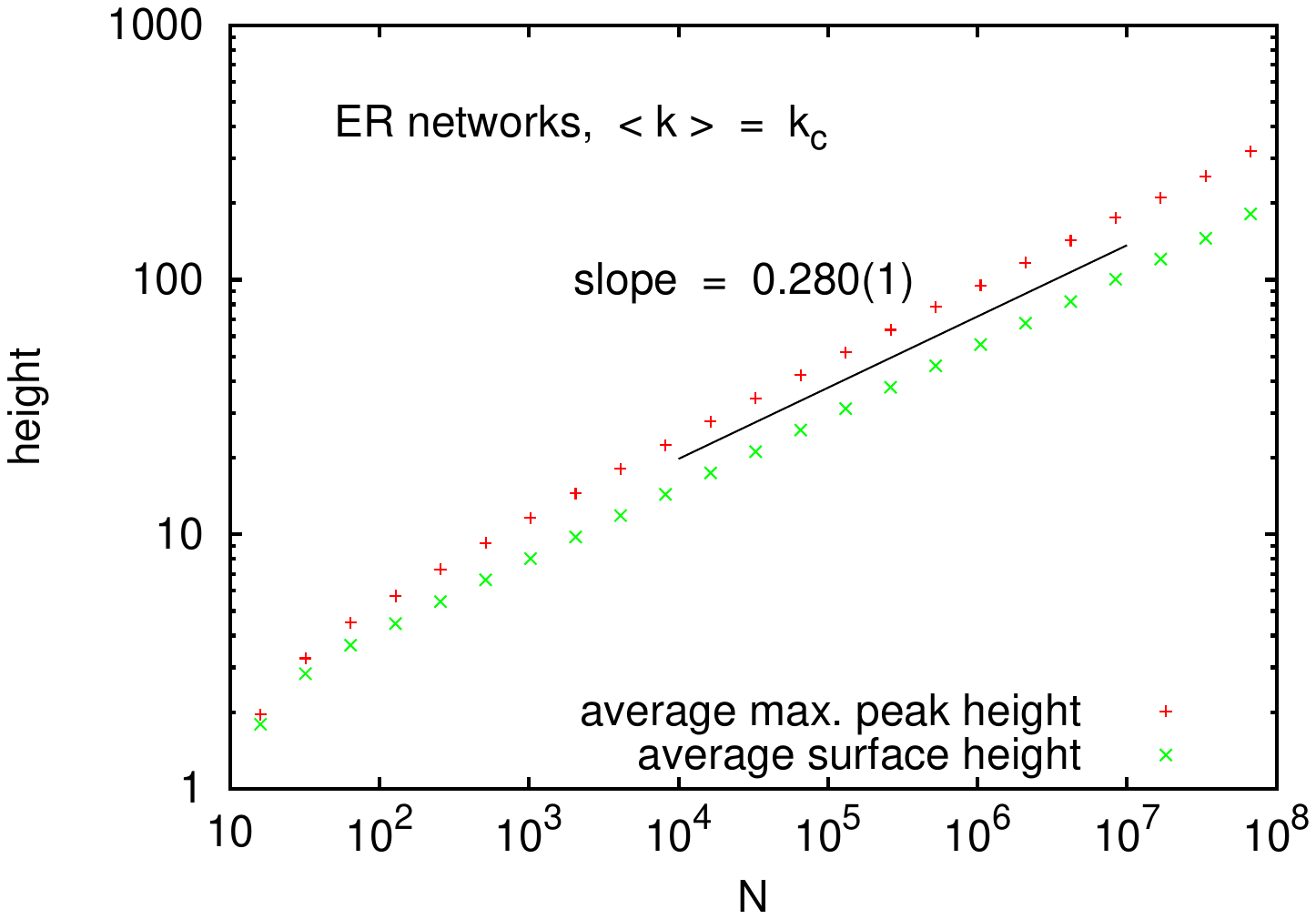}
\includegraphics[scale=0.31]{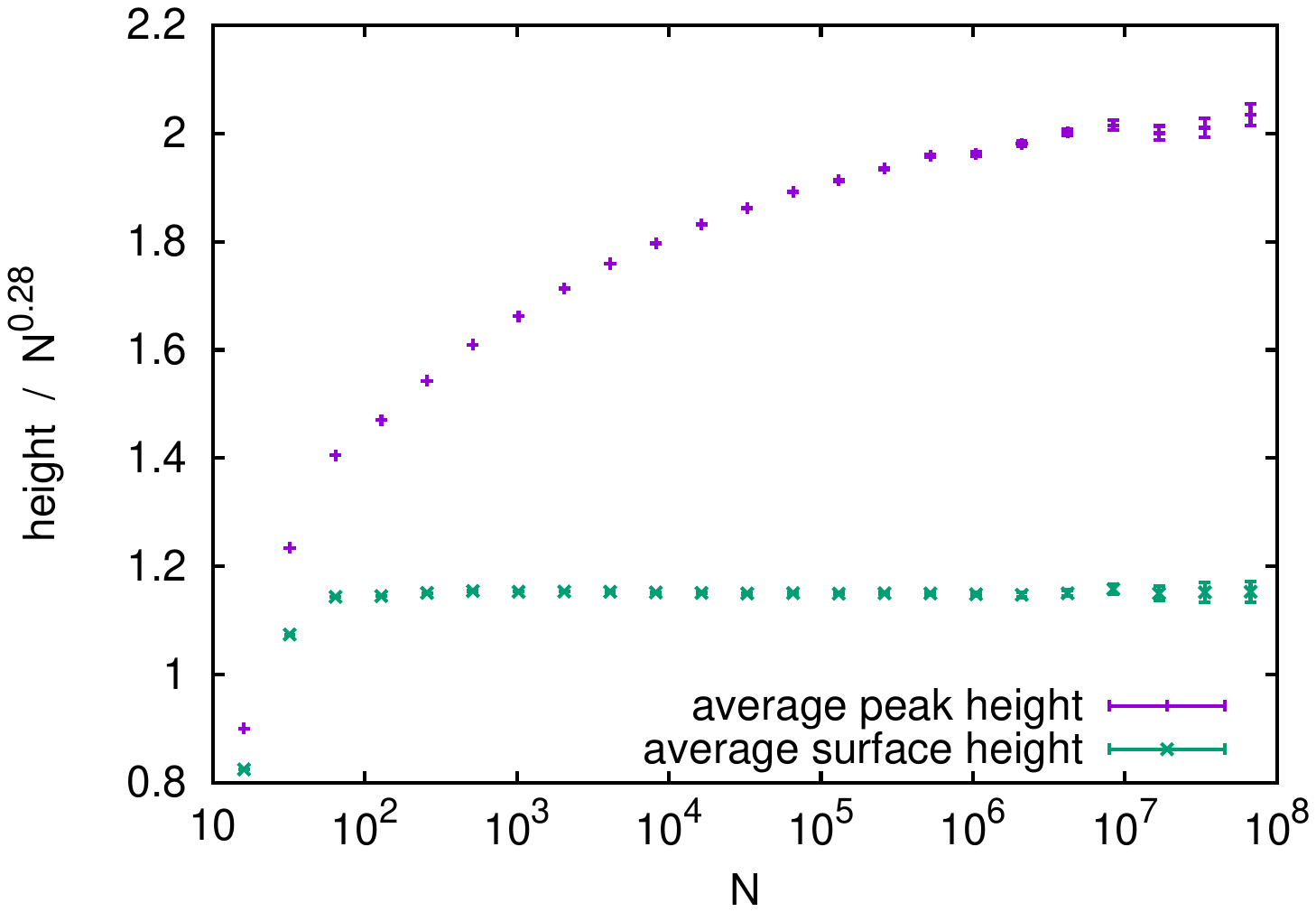}
\caption{(color online) (a) Average ``surface" heights at $z=z_c$ plotted against networks network sizes on a log-log 
plot, and average values of the peaks in each surface. Both curves clearly disagree with the scaling $\propto N^{1/4}$
proposed in \cite{Buldyrev}, but give rather an exponent $0.280(1)$; (b) The same data, but divided by $N^{0.280}$.}
\label{fig.6}
\end{figure}

\begin{figure}[htp]
\includegraphics[scale=0.31]{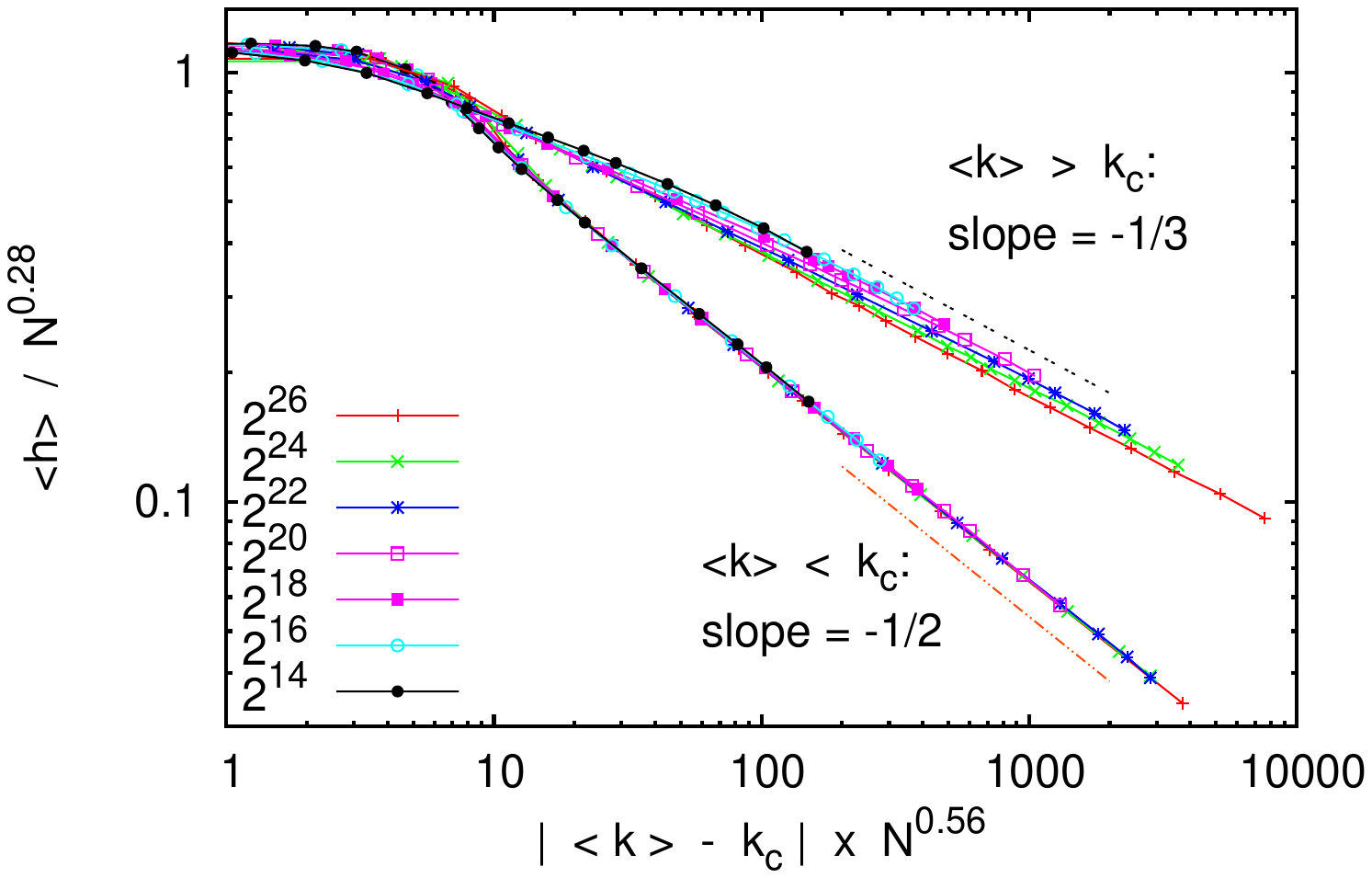}
\caption{(color online) Log-log plots of $\langle h \rangle$ against $|z-z_c|$. The straight lines indicate power 
laws which we claim to hold in the limit $N\to\infty$.}
\label{fig.7}
\end{figure}

The ``wings" of the peaks seen in Fig.~6 are studied more closely in Fig.~8, where we plotted the heights (on 
a log scale) against $\log|z-z_c|$. For the subcritical regime $z<z_c$ we clearly see a power law with exponent
$1/2$, in perfect agreement with \cite{Zhou} (notice, however, that we could not confirm the other scaling laws 
suggested in \cite{Zhou}). We did not try to derive this analytically, but doing this should not be difficult. 
For $z>z_c$ the situation is less clear, but it seems that for large $N$ a power law with exponent $1/3$ emerges, 
even if corrections are large for finite $N$.

\begin{figure}[htp]
\includegraphics[scale=0.31]{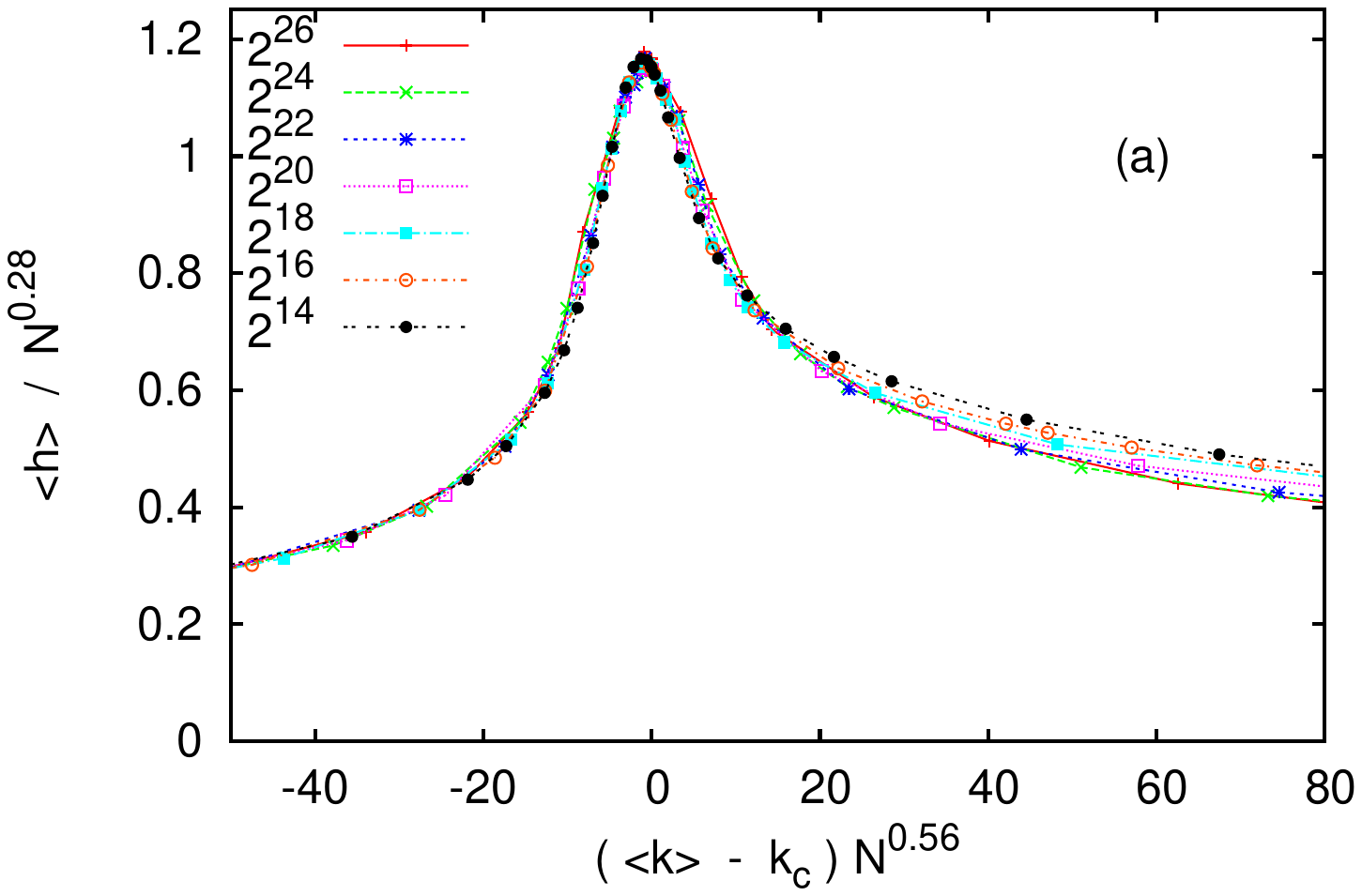}
\includegraphics[scale=0.31]{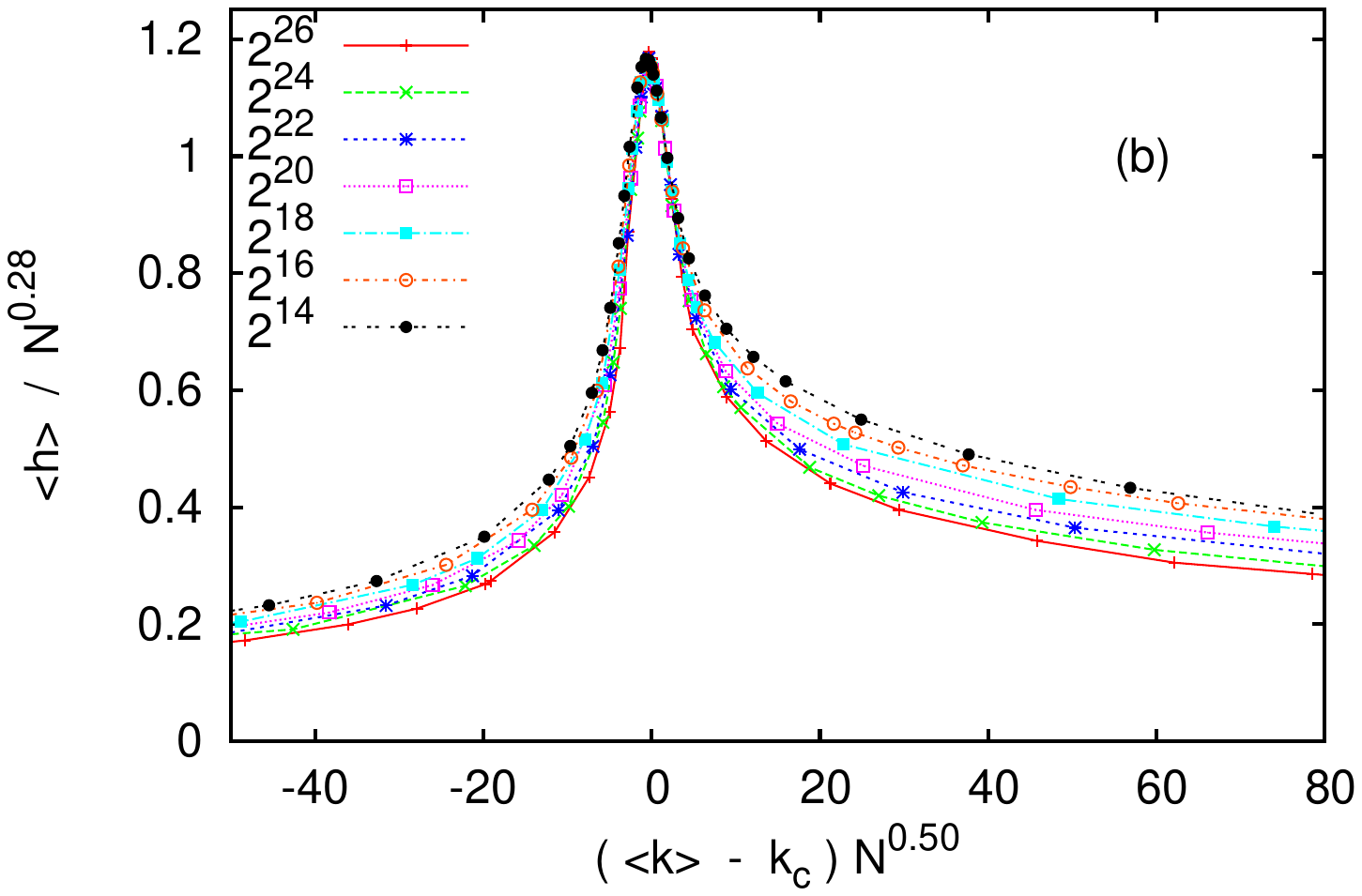}
\caption{(color online) Data collapse plots of re-scaled surface heights $\langle h \rangle /N^\gamma$ against 
$(z-z_c) N^{2\gamma}$ with $2\gamma = 0.56$ (panel a) and against $(z-z_c) N^{0.5}$ (panel b). Neither plot is fully satisfactory, 
but in spite of the bigger overall discrepancies we claim that panel (b) gives the correct collapse in the critical 
region.}
\label{fig.8}
\end{figure}

The two panels of Fig.~9 show finally two attempts to produce a data collapse for the entire critical region. 
In panel (a) we plotted $\langle h \rangle /N^\gamma$ against $(z-z_c) N^{2\gamma}$. This gives us a perfect
data collapse subcritically. In the critical and supercritical regions the plot does not look too bad either,
but a closer inspection shows that it is indeed unacceptable. In contrast, in panel (b) we plotted the data 
against $(z-z_c) N^{0.5}$. This time the collapse is very poor in the wings, but we get a very good collapse
in the critical region -- apart from the same shift of the effective critical point also seen in Fig.~5.
Notice that the variables on the x-axis in Figs. 5 and 9 are the same.

\subsection{2-dimensional Lattices}

We next study 2-d square lattices of size $L\times L$, with $32 \leq L \leq 32768$ and with helical boundary 
conditions \cite{Note4}.
The two sets of links are different (randomly chosen) subsets of nearest neighbor bonds. For each 
of them we include a bond with probability $p$ and exclude it with probability $1-p$. In addition, we can also 
make a fraction $q$ of sites inaccessible. For $p=1$ and $0<q<1$ the red and green links coincide, and we 
simply deal with site percolation with control parameter $q$. On the other hand, if $p<1$ and $q=1$, the sets
of red and green links overlap. Each one of them would define a bond percolation problem, and the viable clusters
are subsets of intersections of red and green percolation clusters. 

For $p<1$ and $q<1$, the single color problems would correspond to mixed site-bond percolation. We
studied also that case, because the problem was originally formulated in \cite{Buldyrev} as a robustness 
problem for network break-down under successive node removal. In that interpretation one needs $p<1$ for 
having a multiplex network, and $q<1$ due to node removal. Actually, however, the problem just becomes more 
complicated by having both $p$ and $q$, without adding any conceptual advantage. If one wants to interpret 
results obtained with $q=1$ (i.e. with an undiluted lattice) in terms of robustness against damage, then 
one simply has to implement the damage not as site removal, but as bond removal.

In the following we shall present results for $q=1$ and varying $p$ (because this is conceptually the simplest), 
and for $p=0.6$ and varying $q$. The latter was done in order to compare with the results of \cite{Son1,Berezin3,Son3}.
It leads to $q_c \approx 0.96$.
In neither case the exact critical point is known. The total number of realizations were in both cases 
between $>10^8$ for the smallest, and several thousand for the largest lattices. 

\begin{figure}[htp]
\includegraphics[scale=0.31]{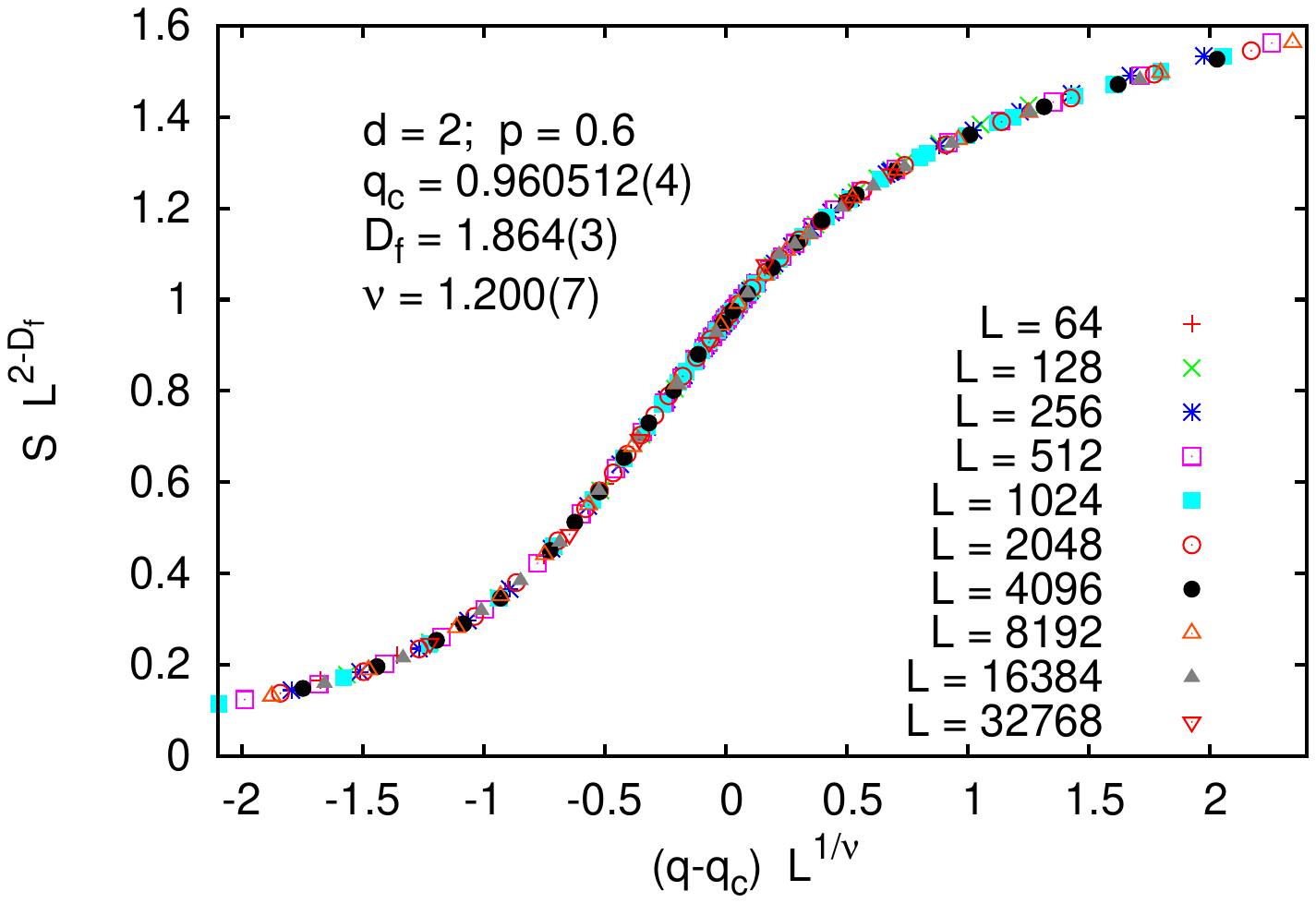}
\caption{(color online). Data collapse plot for the size of the largest viable cluster on 2-d lattices with $p = 0.6$.
    The only perceptible deviation from a perfect collapse is at very large values of $q$, where the data for $L=256$
    and for $L=512$ are systematically too high, due to finite-size corrections.}
\label{fig.9}
\end{figure}

\begin{figure}[htp] 
\includegraphics[scale=0.31]{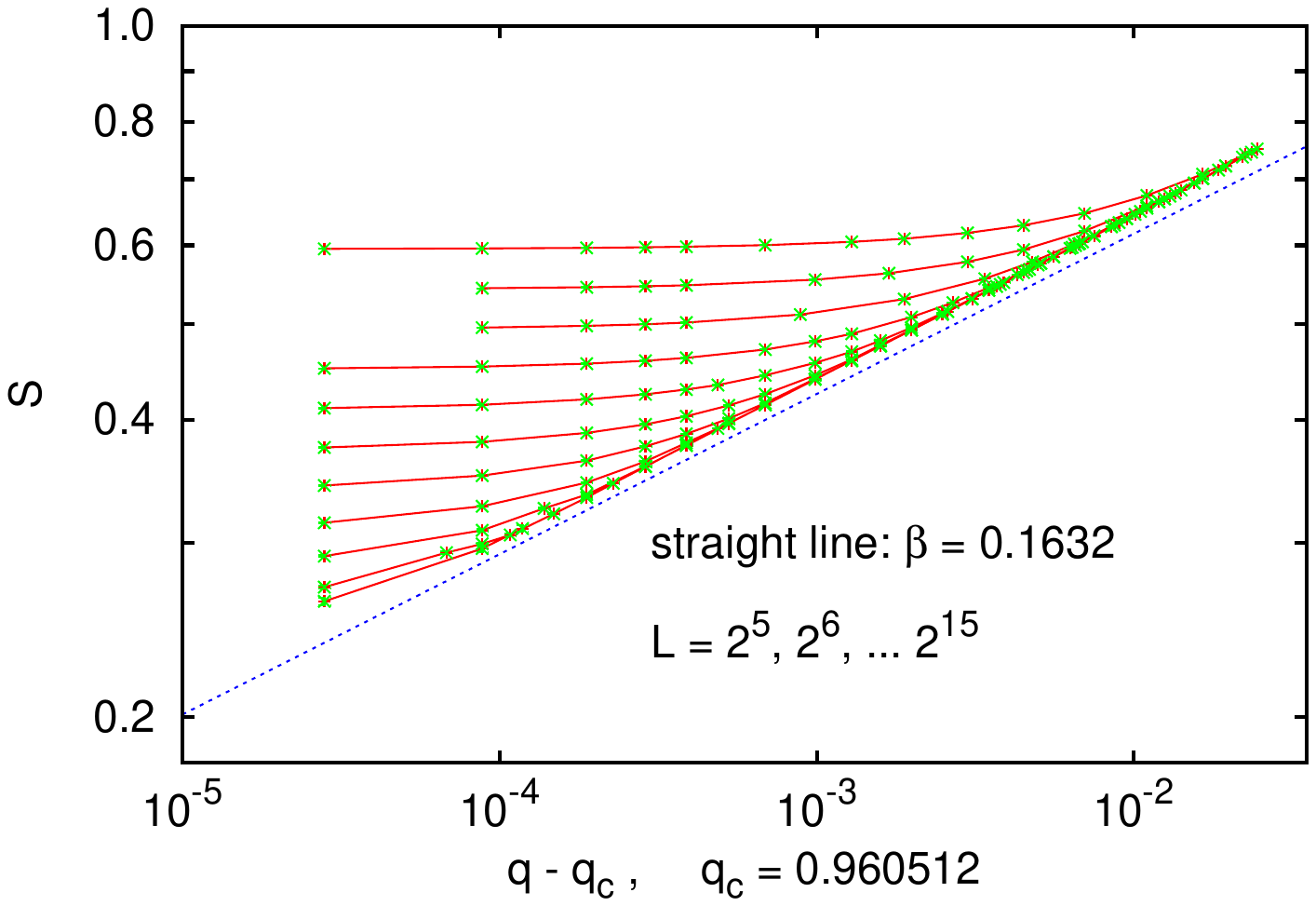}
\caption{(color online). Log-log plot of $S$ against $q-q_c$, demonstrating the power law 
    $S \sim (q-q_c)^\beta.$ }
\label{fig.10}
\end{figure}

\begin{figure}[htp]
\includegraphics[scale=0.31]{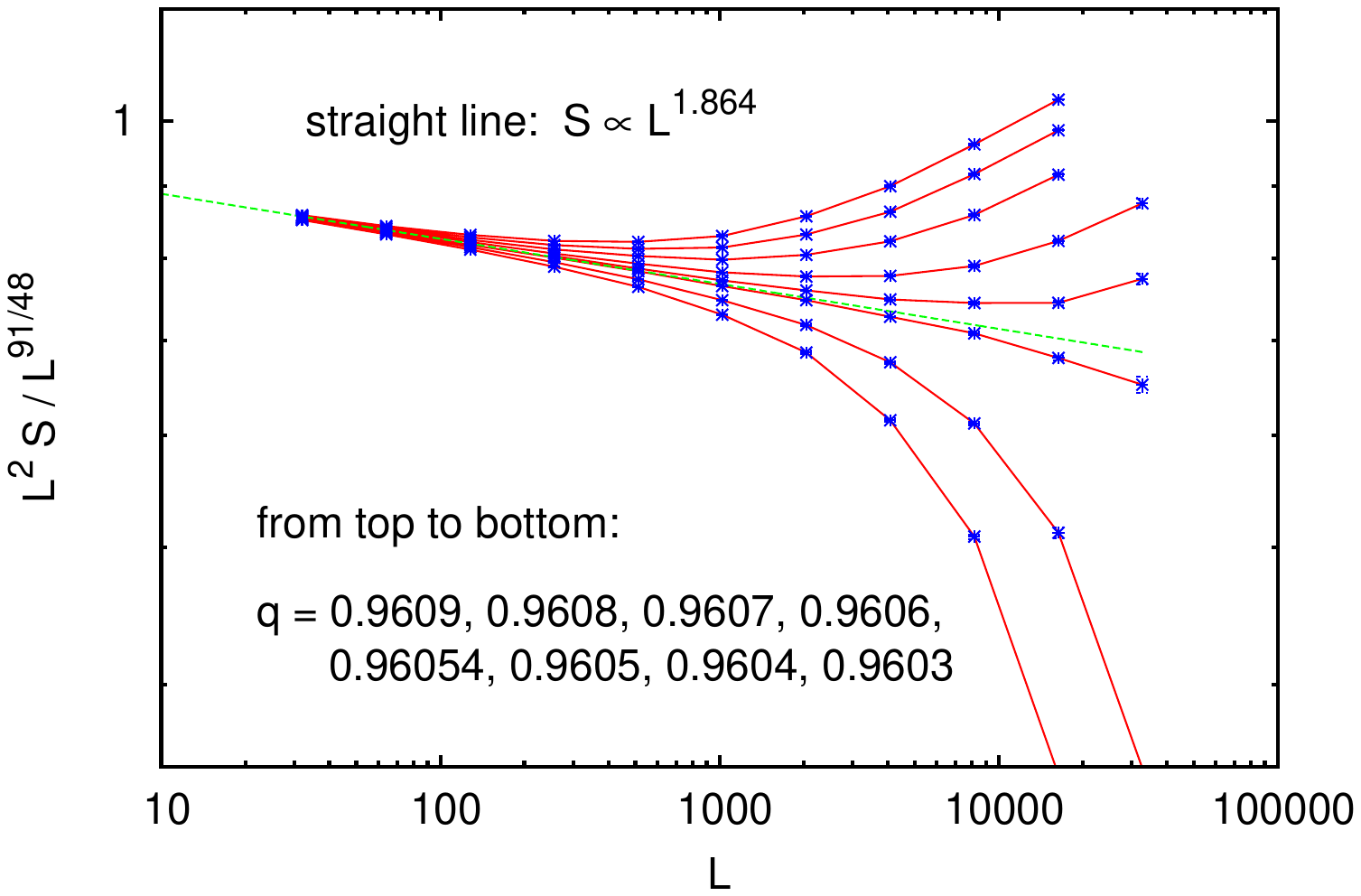}
\caption{(color online) Log-log plot of $L^2 S / L^{91/48}$ against $L$, for eight values of $q$ close to $q_c$.
    The power $91/48 = 1.8958\ldots$ was chosen, because it is the fractal dimension of the incipient giant cluster 
    in critical OP. Thus any deviation of the critical curve from a horizontal line indicates that the present 
    model and OP are in different universality classes.}
\label{fig.11}
\end{figure}

Results for $p=0.6$ are shown in Figs.~10 to 12.
In Fig.~10 we show a data collapse similar to the corresponding plots in \cite{Son1,Berezin3}.
We see a perfect collapse, if we use $q_c = 0.960512(4)$ and $\nu = 1.2$. These values agree roughly with those
proposed in \cite{Son1}, but are very different from those in \cite{Berezin3}. In particular, we would obtain
a very bad collapse (for any value of $\nu$), if we would use $q_c = 0.9609$ as proposed in \cite{Berezin3}. Also, 
we would obtain a very bad collapse (for any value of $q_c$), if we would take for $\nu$ the value 4/3 as in 
ordinary percolation. We should point out that the largest lattice sizes studied in \cite{Berezin3} were 
only $3000\times 3000$.

Supercritical data are plotted as $S$ against $q-q_c$ on a log-log plot. For large $L$ and $q$ not 
too close to $q_c$ we expect a power law 
\be
    S \sim (q-q_c)^\beta.
\ee
This is indeed nicely confirmed, with $\beta = 0.163(2)$. Notice that this is perfectly consistent with the 
scaling relation $D = 2-\beta/\nu$, but is again different from the value for OP. Notice also that the power
law holds only for $q-q_c < 0.01$. For larger distances from the critical point, corrections would be needed.

Finally, we show in Fig.~12 a log-log plot of $S$ against $L$. To make the plot more significant, we 
actually divides  $S$ by the power of $L$ expected for OP, so that the critical curve would be 
horizontal if the present model were in the OP universality class, as claimed in \cite{Berezin3}. The dashed
straight line indicates the power law consistent with the previous two figures.

\begin{figure}[htp]
\includegraphics[scale=0.31]{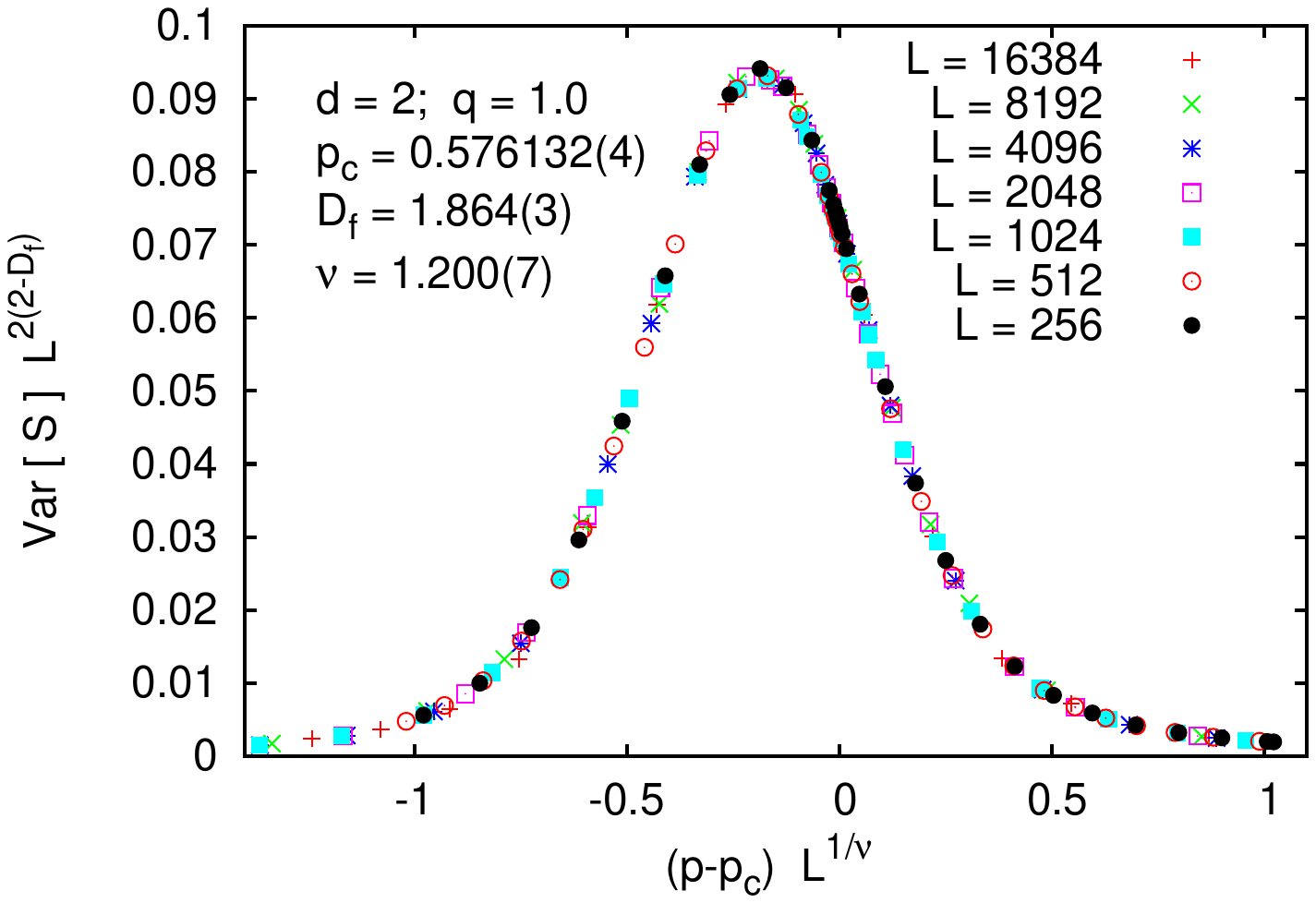}
\caption{(color online). Data collapse plot for the variance of the size of the largest viable cluster on 
    2-d lattices with $q=1.0$, plotted against rescaled values of $p-p_c$.}
\label{fig.12}
\end{figure}

Essentially the same results were obtained when we kept $q=1$ fixed (i.e., no sites were deleted), and $p$ was 
allowed to vary. This time we studied only lattices with sizes up to $16384\times 16384$, but with higher 
statistics. The critical point is now $p_c = 0.576132(5)$. All critical exponents are compatible with these 
given above, and have roughly the same error bars. The main difference with the case $p=0.6$ is that now 
corrections in the far supercritical region do not make the curves in the log-log plot of $S$ against $p-p_c$
turn up (as in Fig.~11), but make then turn slightly down. To avoid duplication we do not show the dame plots as for
$p=0.6$, but we do show a data collapse plot for the variance (``susceptibility") of $S$.

\begin{figure}[htp]
\includegraphics[scale=0.31]{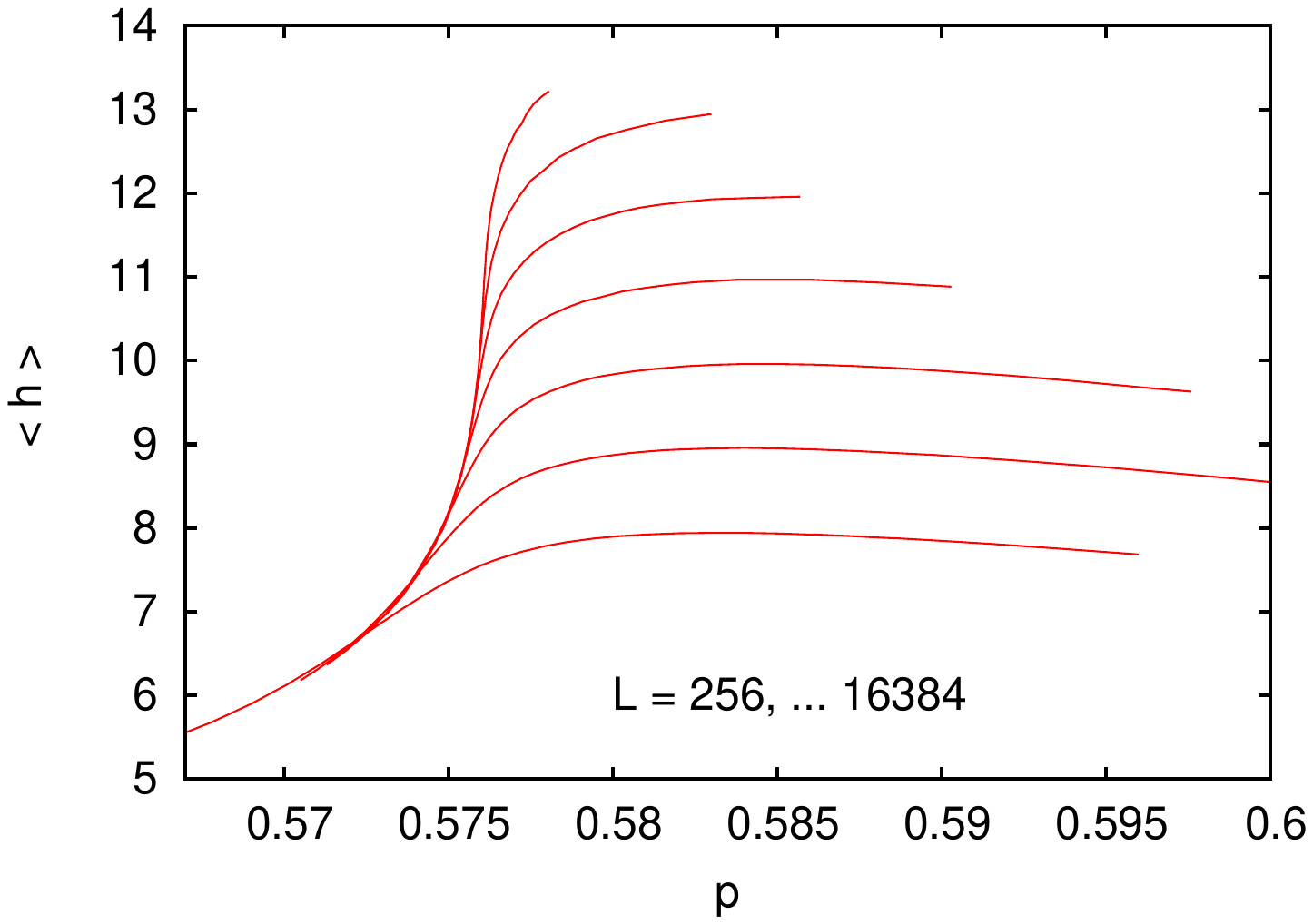}
\caption{(color online) Average SOS landscape heights for 2-d lattices with $q=1$, plotted against $p$.}
\label{fig.13}
\end{figure}

Average landscape heights are shown in Fig.~14. We see a behavior rather different from that seen in Fig.~6 for 
ER graphs. While the latter showed a very sharp peak at the critical point, whose height and sharpness both 
increased with powers of the system size, we have now a very wide bump far in the supercritical region, whose 
width does not seem to depends on $L$ at all, and whose 
height increases logarithmically with $L$. The most conspicuous feature of Fig.~14 is that the slopes
at $p=p_c$ become increasingly steeper with $L$, roughly as 
\be
   \frac{d}{dp} \langle h\rangle \sim L^{0.7} \quad {\rm at} \quad p = p_c.     \label{dhdp}
\ee

In contrast to the ER case, where we could find a decent scaling behavior of $\langle h\rangle$ near the 
critical point, we were unable to find any convincing scaling in the present case.

Before leaving this subsection, we should mention that all scaling laws and critical exponents found in this 
subsection were also found in a very different microscopic realization of the 2-d lattice model discussed in
subsection III.F.

\subsection{3-dimensional Lattices}

For lattices with $d>2$ we made only simulations without site decimations, i.e. we always used $q=1$.

In \cite{Son1}, simulations of very small lattices had suggested that the transition is also for 3-dimensional
lattices continuous and not in the OP universality class. This is important, since it could be argued that 
the transition is continuous for $d=2$ because there the two sets of links necessarily overlap strongly, and 
in that case the transition can be continuous even on ER networks \cite{Parshani}. In three dimensions, this 
overlap is much reduced.

\begin{figure}[htp]
\includegraphics[scale=0.31]{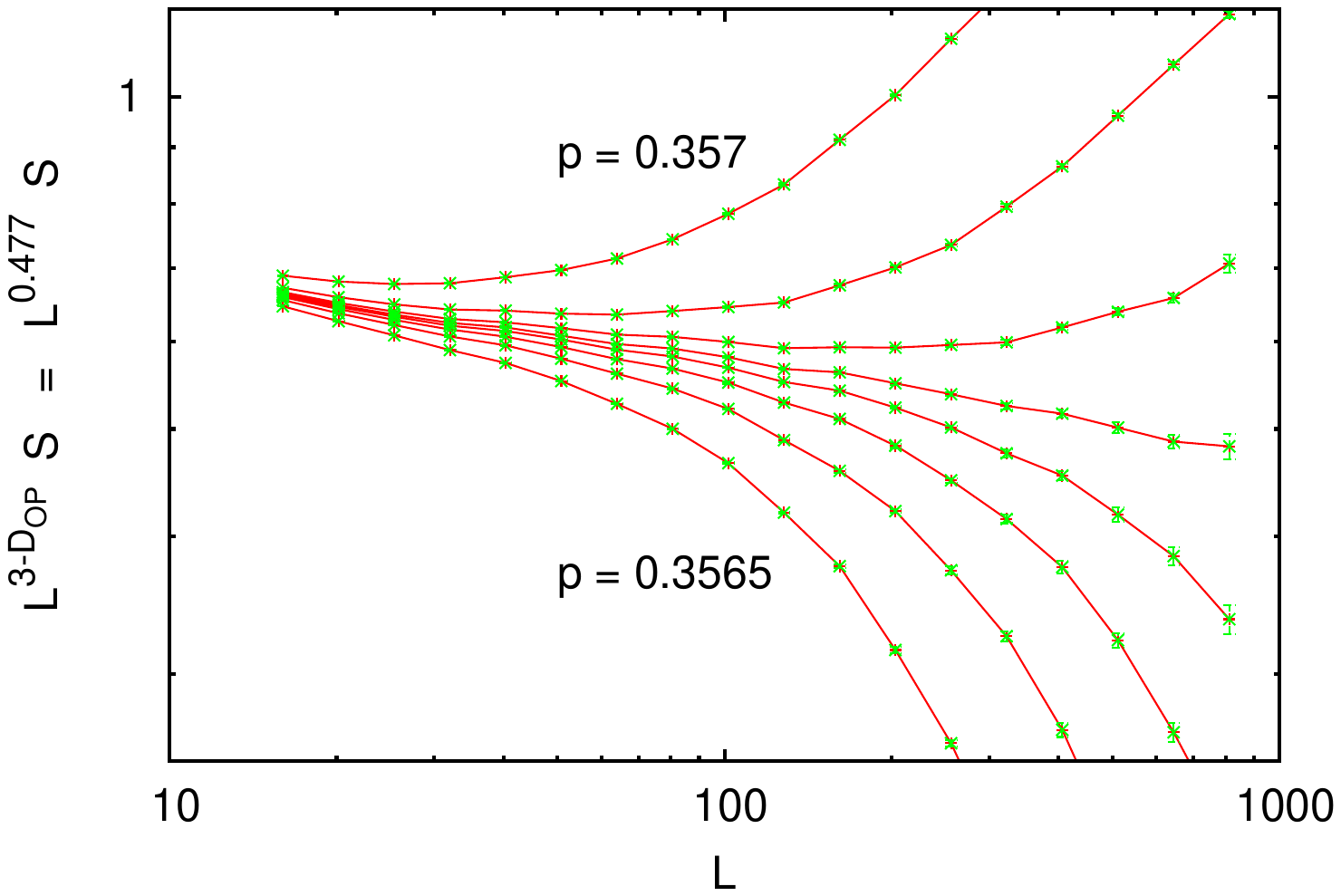}
\caption{(color online) Log-log plot of $L^3 S / L^{D_{OP}}$ against $L$, for eight values of $p$ 
    close to $p_c$. Here, $D_{OP} = 2.523(1)$ is the fractal dimension in three-dimensional OP \cite{Lorenz,Deng}.
    Indeed, since we used helical b.c. where the number $N$ of lattice sites was always a power of 2, some 
    of the lattices were not strictly cubes. In these cases, $L$ was nevertheless defined as $L=N^{1/3}$, even 
    if this was not an integer. The values of $p$ are (from top to bottom) $0.357,0.3568,0.35673, 0.35670, 0.35668,
    0.35665, 0.3566, 0.3565$.}
\label{fig.14}
\end{figure}

Our present simulations show that the transition is indeed continuous, but the deviations from OP scaling are 
much smaller than found in \cite{Son1}. In Fig.~15 we show a log-log plot of rescaled infinite cluster sizes
against $L$ for eight values of $p$ close to $p_c$. Here $L$ is {\it defined} as $L=N^{1/3}$, where $N$ is the 
number of lattice sites. Since we used helical b.c. where $N$ was always a power of 2, the so defined $L$
is not always an integer, and the lattices are not always exact cubes. This has however no noticeable effect
on the scaling. Since we want to compare with OP, where $L^3 S \sim L^{D_{OP}}$ with 
$D_{OP} = 2.523(1)$ \cite{Lorenz,Deng}, we multiply the data with the corresponding power of $L$. 
Thus we expect one curve (the critical one) to be horizontal, if
and only if the present model is in the universality class of OP. We see that the most straight curve 
has a slightly negative slope. This gives $p_c = 0.356707(4)$ and
\be
   D_f = 2.475(8)
\ee
(the estimate of \cite{Son1} was 2.40(1)), which would nominally indicate that the present model is in a 
different universality class. As we shall see, however, there are important corrections to scaling. Thus 
the error in this estimate may be severely underestimated.

\begin{figure}[htp]
\includegraphics[scale=0.31]{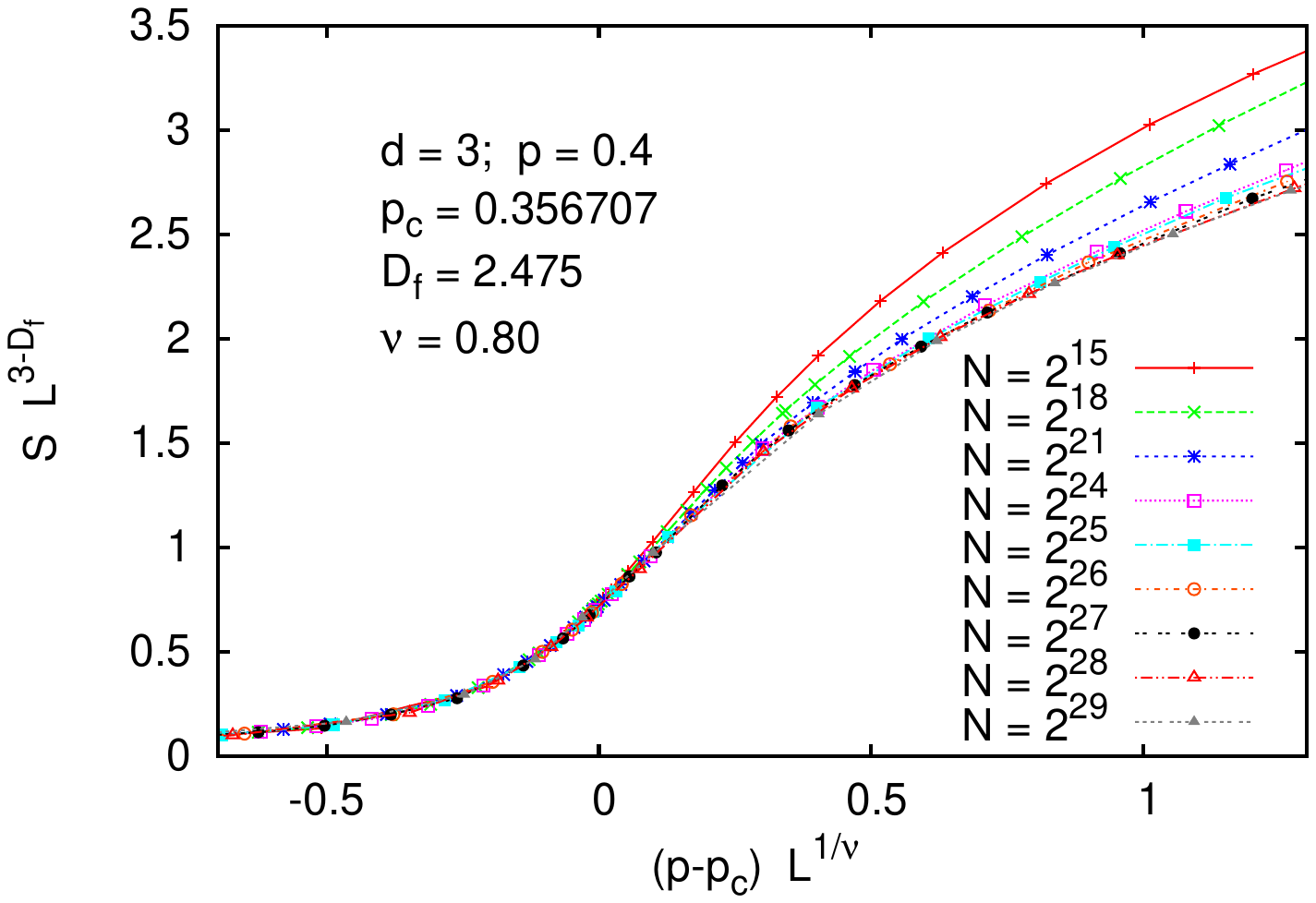}
\caption{(color online). Data collapse plot for the size of the largest viable cluster on 3-d lattices.
    In contrast to the 2-d case we now have huge corrections to scaling in the supercritical region.}
\label{fig.15}
\end{figure}

\begin{figure}[htp]
\includegraphics[scale=0.31]{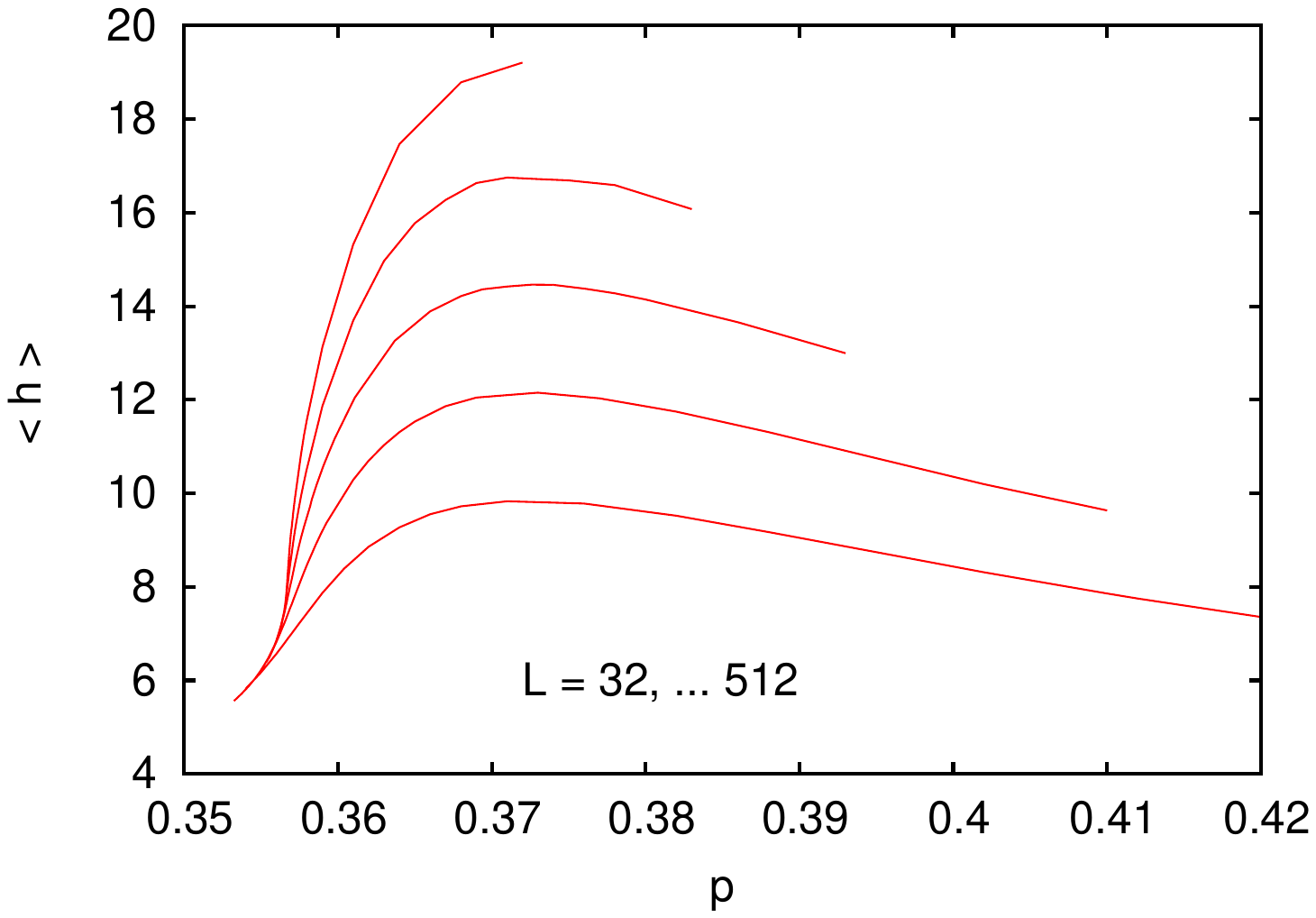}
\caption{(color online) Average SOS landscape heights for 3-d lattices, plotted against $p$.}
\label{fig.16}
\end{figure}

These corrections to scaling are most clearly seen in a data collapse plot, see Fig.~16. In contrast to the 
analogous plot for $d=2$ in Fig.~10, we see now huge deviations in the supercritical region. Nevertheless,
in the critical and under critical regions the collapse seems to be very good. Notice that we used here 
the values for $D_f$ and $p_c$ obtained from the previous plot. For the correlation length exponent we 
obtain $\nu = 0.80(3)$, which is significantly lower that the value $0.8734$ obtained fro OP \cite{Deng}.
Hyper scaling finally gives $\beta = (d-D_f)\nu \approx 0.42$, in agreement with OP. In summary it seems 
that corrections to scaling were severely misjudged in \cite{Son1}, and we cannot exclude that the model
is in the OP universality class for $d=3$. 

A reason for it to be {\it not} in the OP universality class is, of course, the divergence of the surface 
height when $L\to \infty$. This is clearly seen from Fig.~17, where we find the same logarithmic increase 
as in $d=2$, and a power law for the slope at $p=p_c$ as in Eq.~(\ref{dhdp}), but with exponent $\approx 
0.82$.  

\subsection{4-dimensional Lattices}

\begin{figure}[htp]
\includegraphics[scale=0.31]{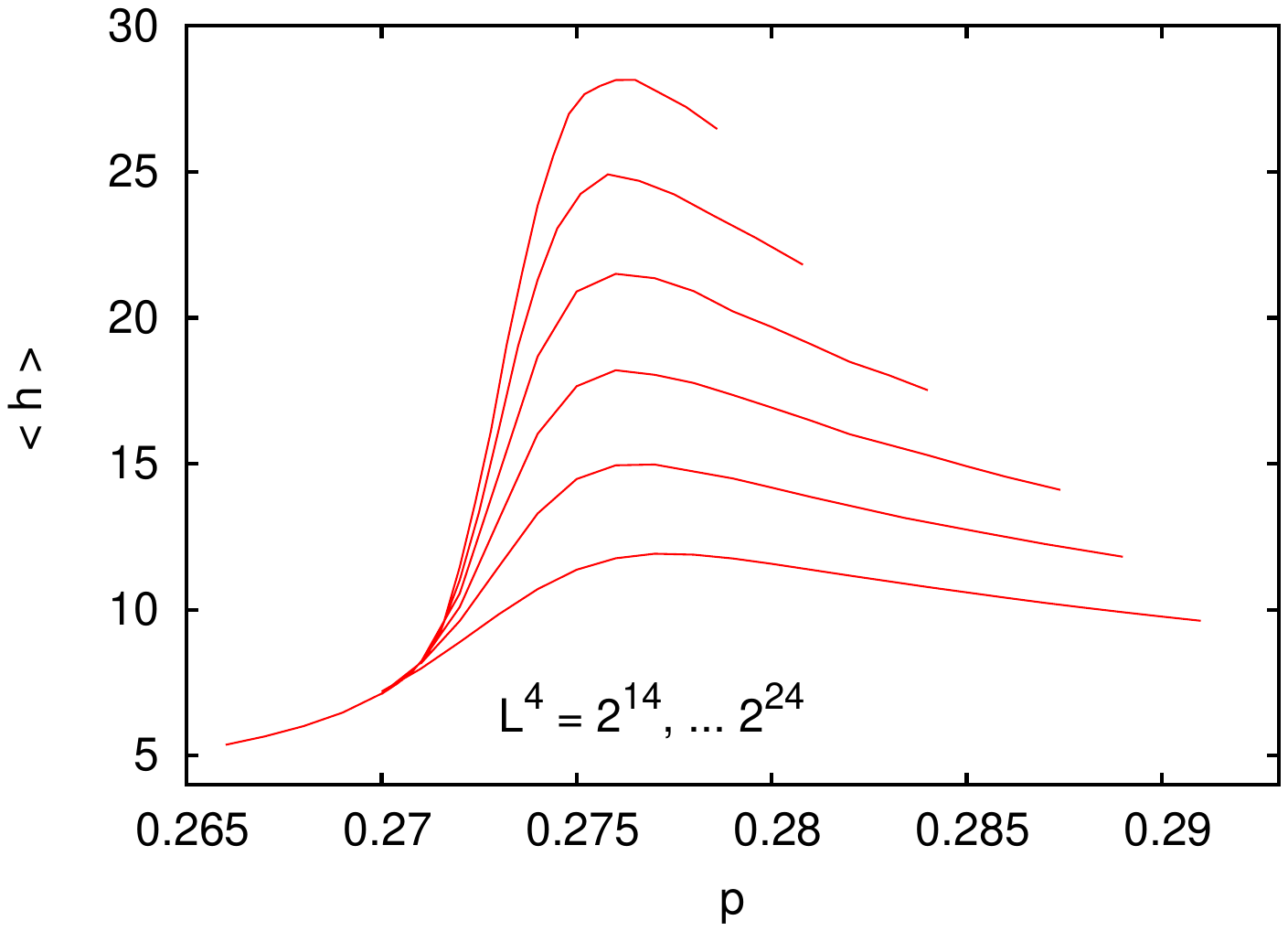}
\caption{(color online) Average SOS landscape heights for 4-d lattices, plotted against $p$.}
\label{fig.17}
\end{figure}

As a first result we show the average surface heights, see Fig.~18. Their behavior is very similar to those
in two and three dimensions: For any $N = L^4$ their maxima occur in the supercritical region, and their
heights increase logarithmically with $L$. Also, the slopes $d\langle h\rangle/dp$ diverge for $L\to\infty$
with a power law, again with power slightly smaller than one (our best estimate is $\approx 0.9$).

\begin{figure}[htp]
\includegraphics[scale=0.31]{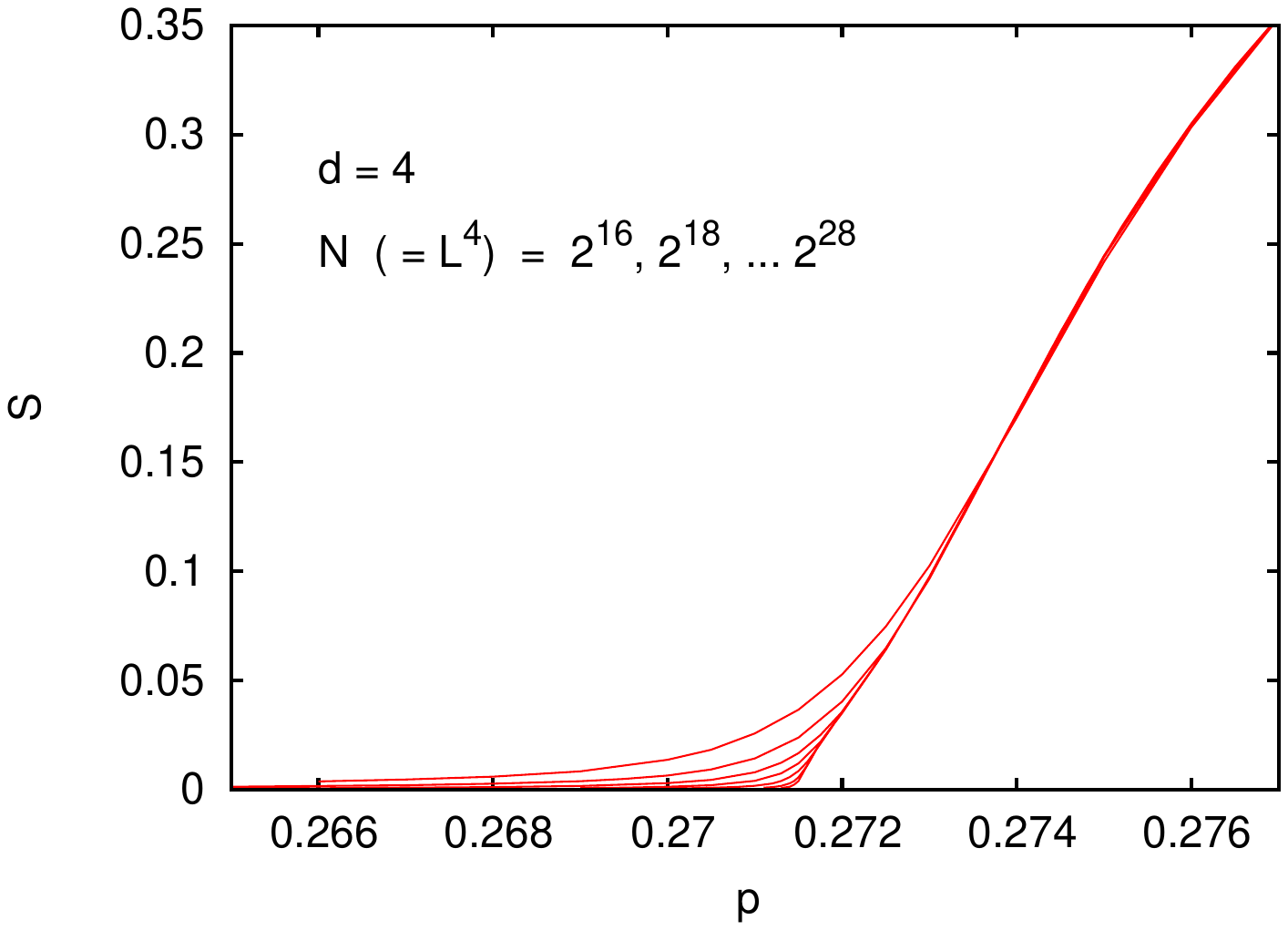}
\caption{(color online) Average density of the largest viable cluster on 4-d lattices with $N$ sites, plotted
    against $p$. The sharpness of the kink at $p\approx 0.2715$ increases with $N$.}
\label{fig.18}
\end{figure}

\begin{figure}[htp]
\includegraphics[scale=0.31]{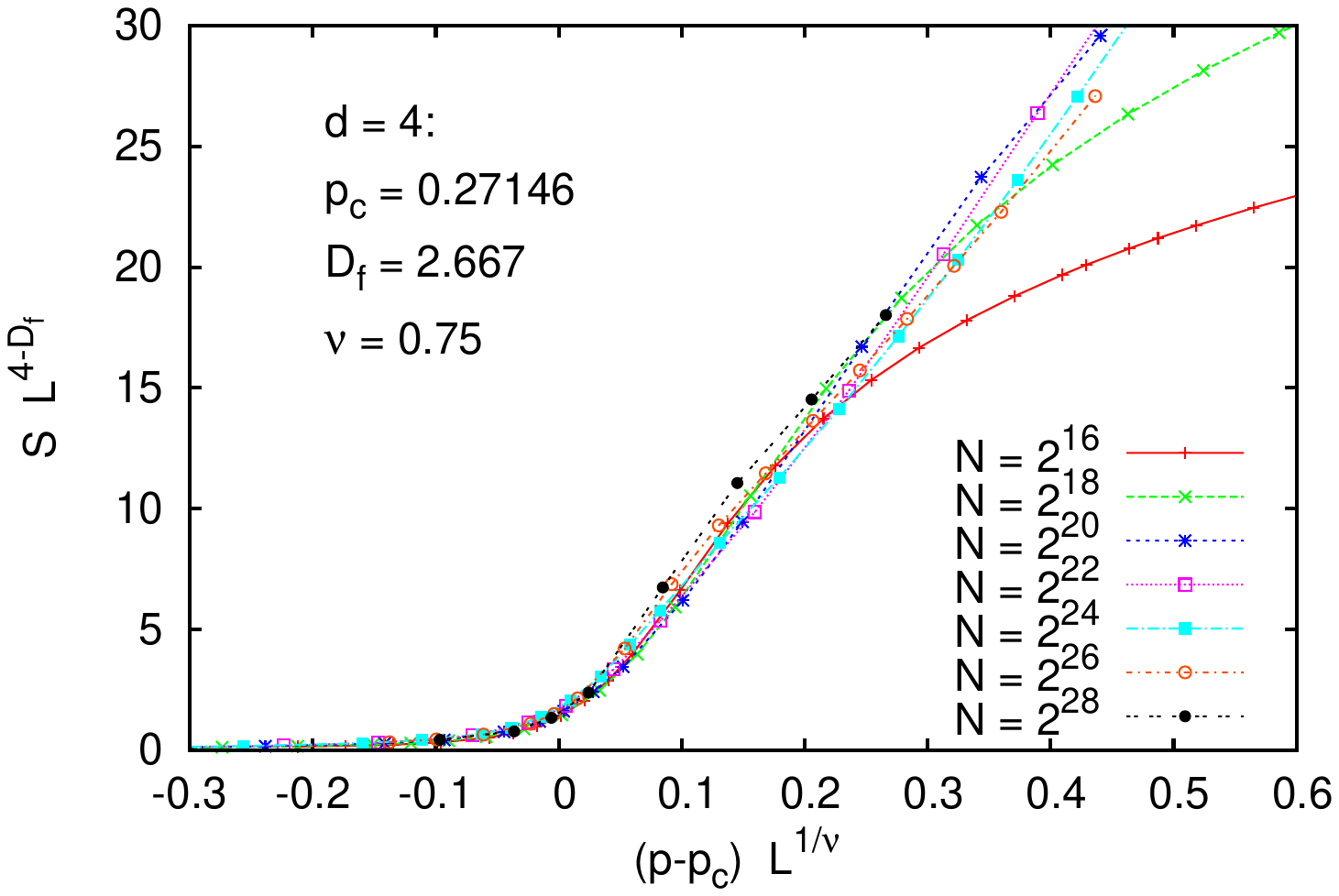}
\caption{(color online) The same data as in Fig.~19, but plotted such that the collapse onto a single 
    curve near the critical point.}
\label{fig.19}
\end{figure}

In contrast to this, the behavior of the order parameter is completely different. In Fig.~19 we show 
plots of $S$ versus $p$, for lattice sizes with $N = 2^{16}, 2^{18}, \ldots 2^{28}$, corresponding to 
$L =16, \ldots 128$. We see clearly a continuous transition, but with order parameter equal to (or at 
least very close to) $\beta=1$. This would indicate that $d=4$ is the upper critical dimension, in 
striking contrast to OP, where $d_u = 6$. An attempt to collapse the data according to standard FSS
leads to Fig.~20, and to $D_f \approx 8/3$ and $\nu \approx 3/4$. These would not correspond to any known 
mean field theory, but we should be aware that we should expect logarithmic corrections, if indeed 
$d_u = 4$. The poor quality of the collapse might indicate that such corrections are indeed present,
but with our present data we have no chance to estimate them in detail.

\begin{figure}[htp]
\includegraphics[scale=0.31]{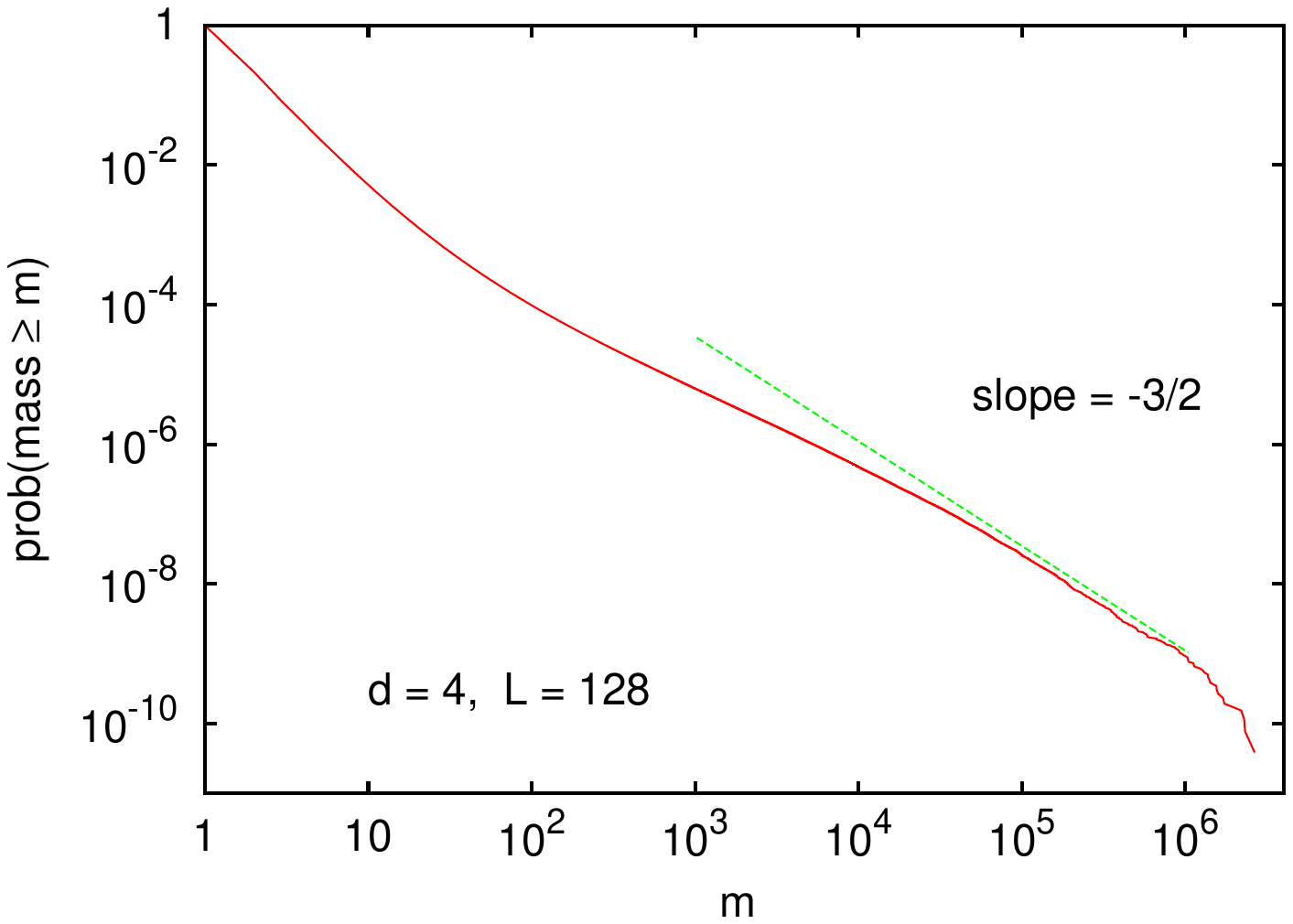}
\caption{(color online) Log-log plot of the integrated mass distribution of 4-d viable clusters at $L=128$
   and $p=0.27145$. The straight line has slope $-3/2$ as predicted from standard scaling laws, accepting the 
   exponents found in Fig.~20.}
\label{fig.20}
\end{figure}

Finally, we show in Fig.~21 the integrated mass distribution on a large lattice for $p\approx p_c$.
If usual FSS would hold, we would expect a power law 
\be
   p(m) \sim m^{1-\tau}                 \label{tau}
\ee
with 
\be
    \tau = 1+d/D_f \;.
\ee
Inserting here our numerical result $D_f = 8/3$ gives $\tau = 2.5$. The data show a very rough power law, 
with large deviations at intermediate mass values, but the distribution agrees surprisingly well 
with Eq.~(\ref{tau}) for large masses.

\subsection{5-dimensional Lattices}

\begin{figure}[htp]
\includegraphics[scale=0.31]{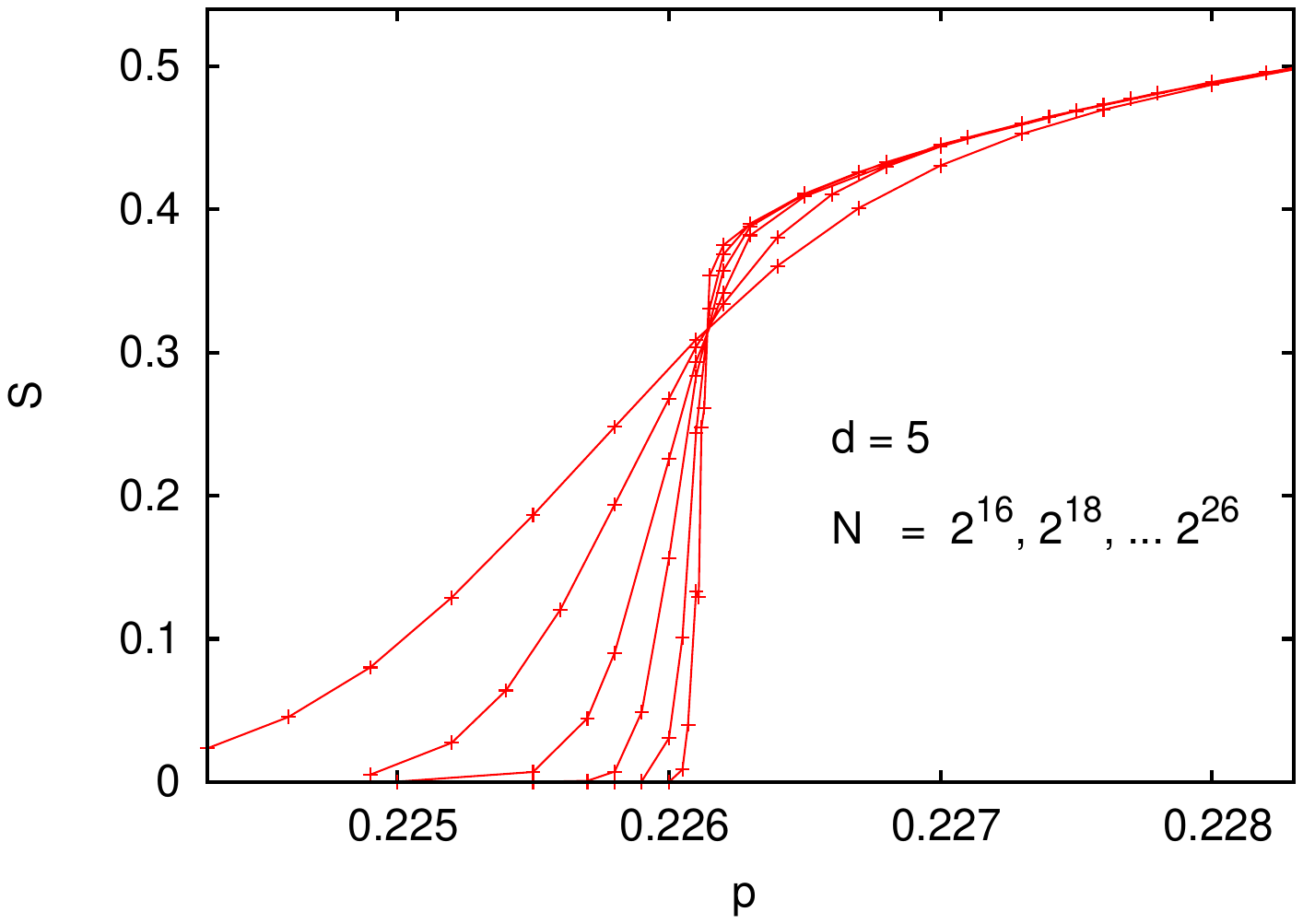}
\caption{(color online) Average density $S$ of the largest viable cluster on 5-d lattices with $N$ sites, 
    plotted against $p$. Notice that we averaged here over all runs (even if they did not have a giant 
    viable cluster), while the averages in subsection III.A (e.g. Fig. 2) were done only over runs that 
    did have a giant cluster.} 
\label{fig.21}
\end{figure}

If the present problem represents indeed a new universality class 
with upper critical dimension $d_u = 4$, then the behavior in $d=5$ should be essentially the same as 
for $d\to\infty$, and in particular as on ER graphs. To verify this we show first (in Fig.~22) the order 
parameter $S$ (the density of the largest viable cluster) as a function of $p$. We see that the curves 
for different sizes $N$ intersect indeed in a single point, with $p_c = 0.22614(1)$ and $S_c = 0.31(1)$.
This jump of $S$ is a clear sign of a discontinuous transition, although the rise of $S$ for $p>p_c$
indicates that the transition is also, as for ER networks, hybrid.

\begin{figure}[htp]
\includegraphics[scale=0.31]{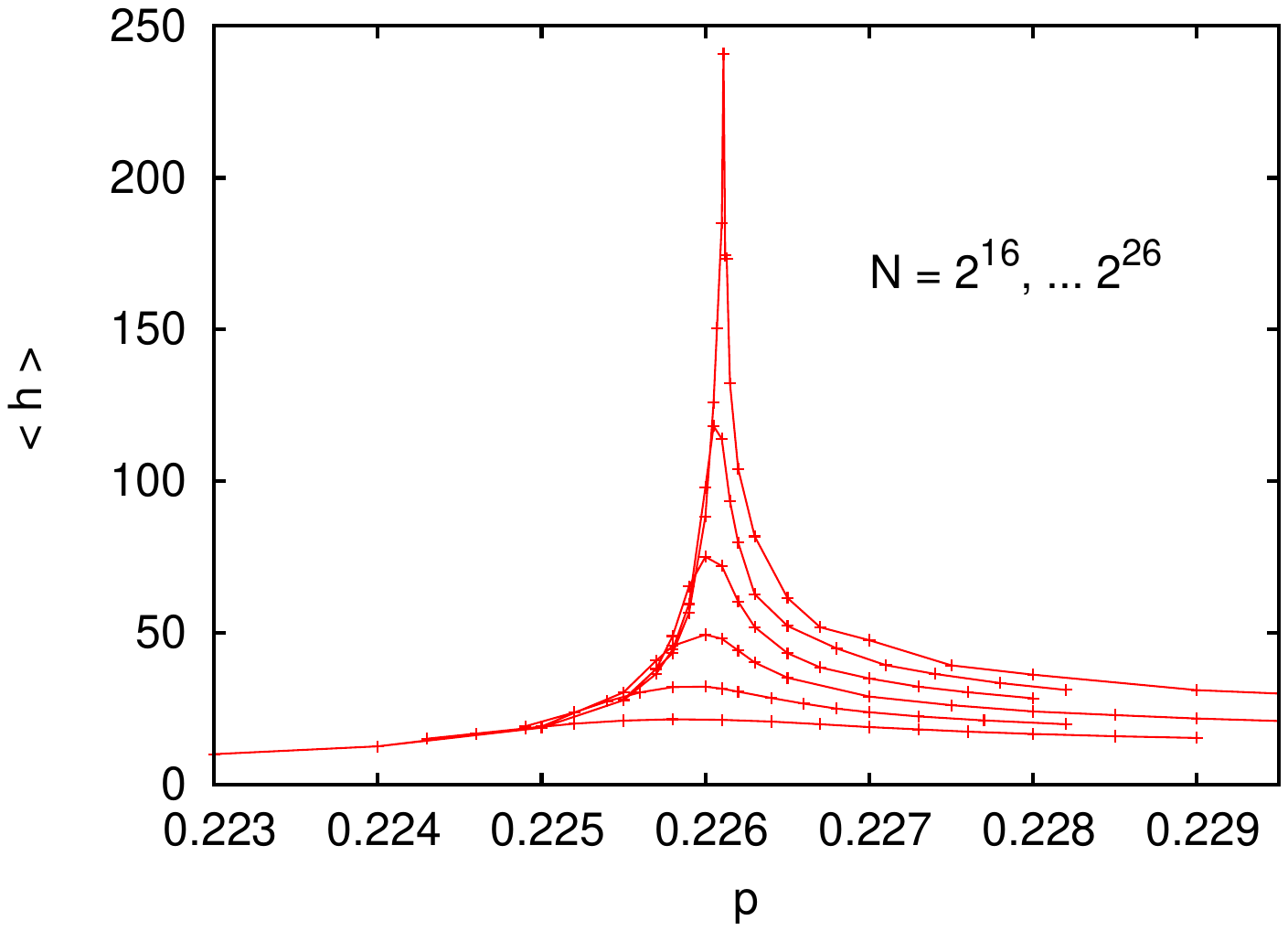}
\caption{(color online) Average SOS landscape heights for 5-d lattices, plotted against $p$.}
\label{fig.22}
\end{figure}

The same conclusion is reached by looking at surface heights (Fig.~23), which display the same sharp 
peak as for ER networks. Also, the heights of the peaks increase with roughly the same power of the 
system size ($h_{\rm max} \sim N^{0.30(2)}$), not logarithmically as on lattices with $d\leq 4$.

\begin{figure}[htp]
\includegraphics[scale=0.31]{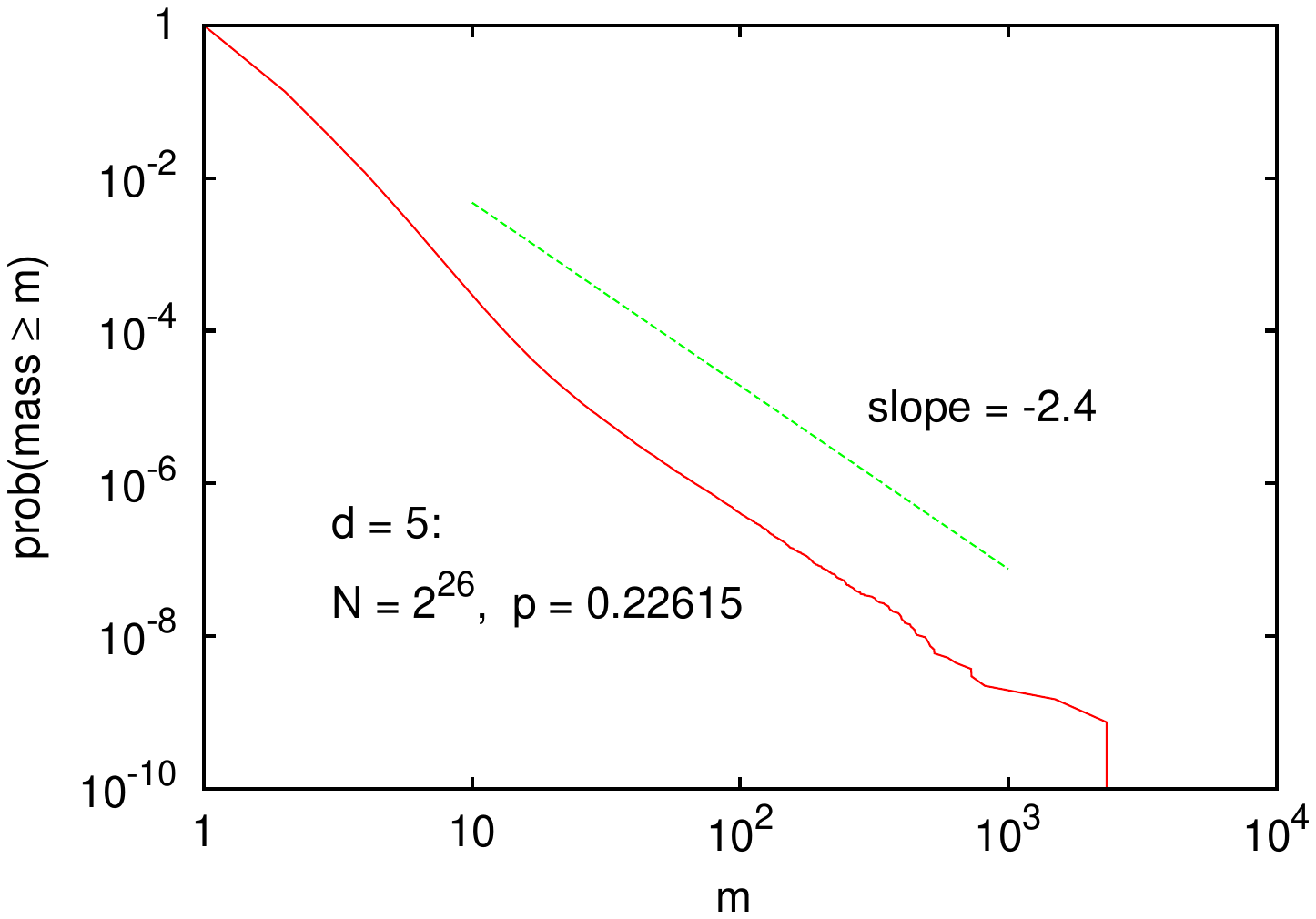}
\caption{(color online) Log-log plot of the integrated mass distribution of finite 5-d viable clusters at 
   $N = 2^{26} \; (L\approx 37)$ and $p = 0.22615$. The straight line indicates the scaling of finite but
   large viable clusters.}
\label{fig.23}
\end{figure}

On the other hand, the statistics of finite viable clusters is markedly different from that for ER
networks. There, apart from the giant viable cluster all other clusters are of size one in the 
$N\to\infty$ limit \cite{Buldyrev}, and for the system sizes studied in the present work most of them 
have sizes one or two. On five-dimensional lattices, however, the distribution of finite viable 
clusters is non-trivial. The average mass of the second-largest clusters peaks for $p_c-p \sim L^{-2}$,
and the peak height increases roughly as $L^2$ (both of these exponents are very crude estimates).
Moreover, the mass distribution of the finite clusters shows rough scaling, 
$ p(m) \sim m^{-2.4} $ at $p=p_c$ (see Fig.~24).

\subsection{Cross-over to Mean Field Theory via Long Range Links on 2-d Lattices}

\subsubsection{General Remarks: Long Range Connectivity  {\it vs.} Long Range Dependency}

In general, there are two main ways how the anomalous behavior of critical phenomena seen on low
dimensional lattices can cross over to mean field theory. One is by increasing the dimension,
the second is by making the theory more and more non-local. In the present model we found that 
the transition via dimension increase is very non-standard, by showing first a transition to a 
mean field {\it critical} point (at $d=4$), which is then replaced by a first order transition at
$d>4$. In view of this, it is of interest to check what mechanism(s) are at play when one makes 
the model more and more non-local.

In the interpretation as multiplex networks (where any interdependent node pairs are joined into
a single node) adopted in this paper, the most natural way to do this is by increasing the lengths
of the (``connectivity") links. In analogy with standard critical phenomena one would expect that 
the model stays in the same universality class as long as the characteristic length of the bonds are 
finite, and crosses over to mean field theory only when this length becomes infinite, e.g. by having
power-behaved link length distributions (see, e.g., \cite{Grass2013,Grass2013b}). One cannot rule 
out, however, that there exists a tricritical point at some finite length. 

In the original
interpretation of \cite{Buldyrev} as a model with distinct connectivity and dependency links, one 
can make either of them long ranged, or both.
In \cite{Li-Bashan,Berezin2,Danziger} it is assumed that nodes are located on 2-d lattices with nearest
neighbor connectivity links, while dependency links become long ranged. This is presumably not very 
realistic in most applications:
Since dependency links are much more crucial than connectivity links, one would assume that they are 
the shorter ones in any realistic natural network. In the following we shall just make some comments
on the model of \cite{Li-Bashan,Berezin2,Danziger}, and leave the more natural model of multiplex 
networks with long range connectivity links to a future publication.

\subsubsection{Random Dependency Links}

As in \cite{Li-Bashan}, we first discuss the case where the dependency links are completely random
(while, as we said, the connectivity links are the nearest neighbor bonds). The transition is obtained 
by varying the site
occupancy $q$ of one of the two dependency partners (i.e., any pair of dependency partners is present
with probability $q$), and keeping all dependency links (i.e., $p=1$). 
This is the $r\to\infty$ limit of the 
model discussed in the next sub-subsection. Following cascades, it was shown in \cite{Li-Bashan} that the 
transition is first order in this limit, and closed formulas were given for the threshold 
$q_c$ and for the density of the viable cluster at threshold.

As in the case of random locally tree-like graphs \cite{Son1,Son2}, these formulas are indeed obtained 
more easily from consistency considerations without reference to any cascade dynamics. 
Let us denote by $S_{\rm OP}(p)$ the order parameter (i.e. the density of the infinite cluster)
in ordinary site percolation with site occupancy $p$ (the subscript ``OP" stands for `ordinary
percolation'). Let us now consider the percolation of one type of nodes, say of nodes of type A. 
Since dependencies are completely random, they are not correlated with the 
(non-trivial) order parameter fluctuations, and they just mean that a finite density of A nodes 
will be killed by not having a viable dependency partner. This fraction is exactly the same as the
fraction of B nodes that are killed because they have no viable A partner. As a result, 
starting with occupancy $p$ in ordinary site percolation leads to the same order parameter as 
starting with occupancy $q$ in the dependency model, provided that \cite{Li-Bashan}
\be
    q = p \times p/S_{\rm OP}(p) \;,  \label{pq}
\ee
and $S_{\rm OP}(p)$ is also the density of the viable cluster when starting with $q$. Plotting 
the r.h.s. of Eq.~(\ref{pq}) versus $p$ gives a convex function (see Fig.~25), the minimum of 
which is the critical value of $q$,

\begin{figure}[htp]
\includegraphics[scale=0.31]{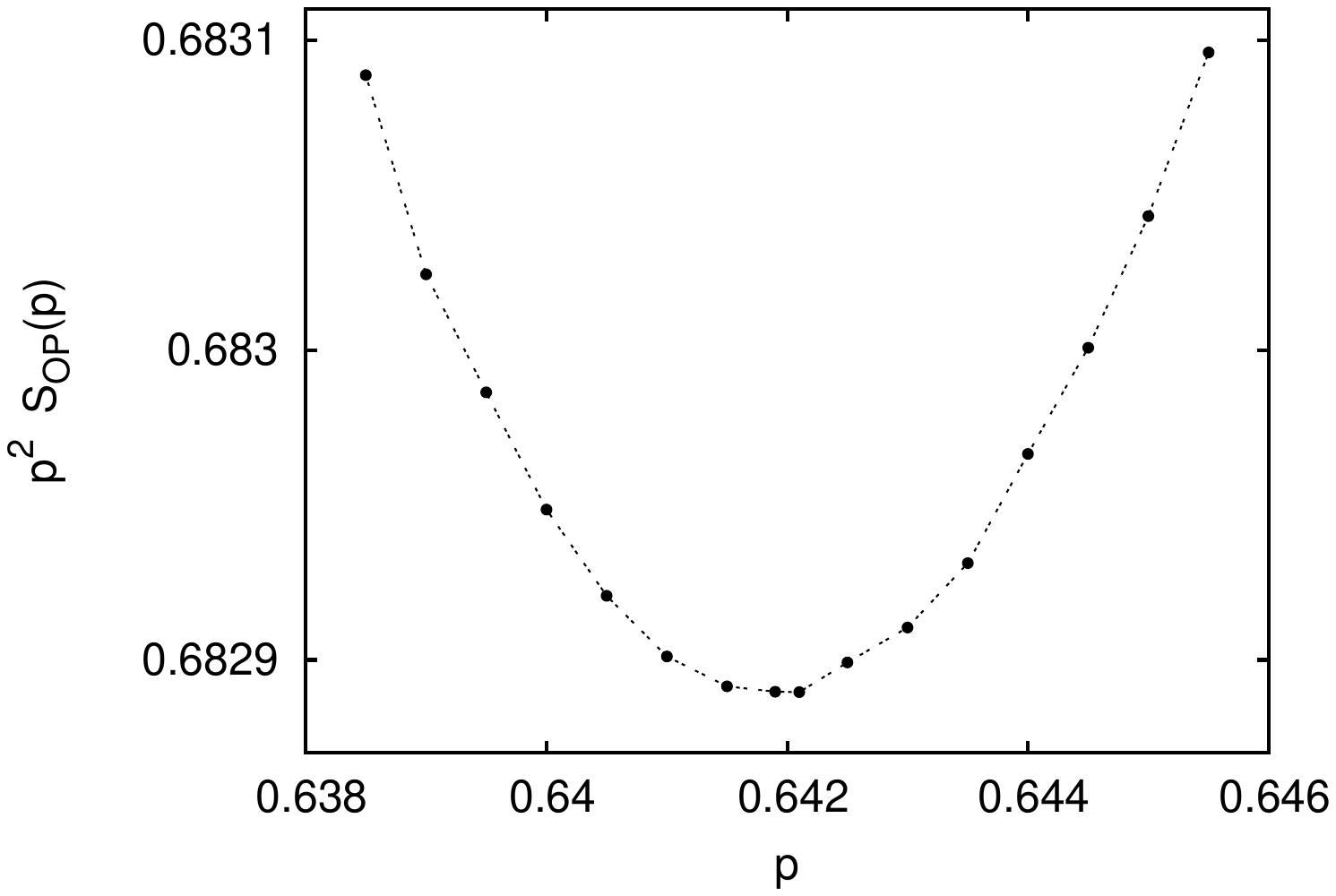}
\caption{(color online) Plot of $p^2/S_{\rm OP}(p)$ versus $p$. The points are most precise near 
   the minimum of the curve, where each point is obtained from $\approx 10^3$ lattices of size 
   $32768\times 32768$. The statistical errors of these points are smaller than their sizes.}
\label{fig.24}
\end{figure}

\be
    q_c = \min_p p^2/S_{\rm OP}(p) = 0.682892(5)
\ee
with 
\be
    p^* = \arg \min_p p^2/S_{\rm OP}(p) = 0.6418(2),
\ee
and $S_c \equiv S_{\rm OP}(p^*) = 0.6031(2)$ is the density of the giant viable cluster at threshold. 
The numerical values are in good agreement with the less precise estimates of \cite{Li-Bashan}.

\begin{figure}[htp]
\includegraphics[scale=0.31]{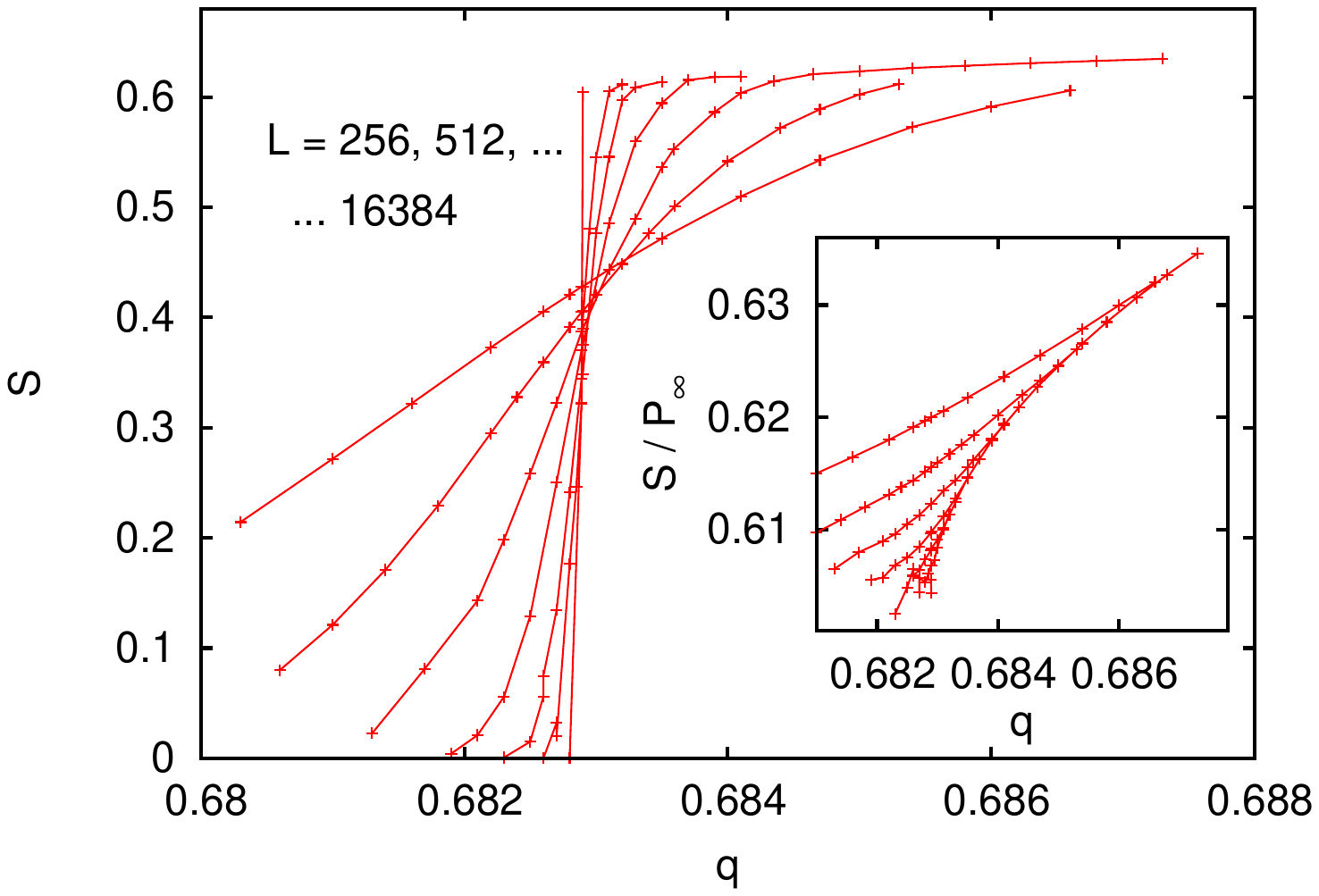}
\caption{(color online) Average density $S$ of the largest viable cluster on 2-d lattices with 
    random dependency links and dependency links between nearest neighbors, plotted against $q$. 
    In the main figure, averages are taken over all runs (even they did not lead to a giant viable 
    cluster, while the inset shows averages restricted to runs with giant viable clusters.}
\label{fig.25}
\end{figure}

The existence of a first order transition with these parameter values was also confirmed by direct
simulations. In Fig.~26 (main plot) we show the order parameter (the density of the largest cluster, 
averaged over {\it all} runs) plotted against $q$. Densities obtained by averaging only over 
those runs which did lead to a giant viable cluster are shown in the inset. As in the ER case,
it was trivial to distinguish between runs with and without giant clusters, because all non-giant
clusters had sizes 1 or 2 (except on very small lattices, where also clusters of size 3 were seen).
Figure 26 confirms not only the values of $q_c$ and $S_c$, but it also shows that the transition 
is hybrid, because the slope at threshold is infinite. The latter is a direct consequence of the 
fact that the minimum in Fig.~25 is quadratic.

\begin{figure}[htp]
\includegraphics[scale=0.31]{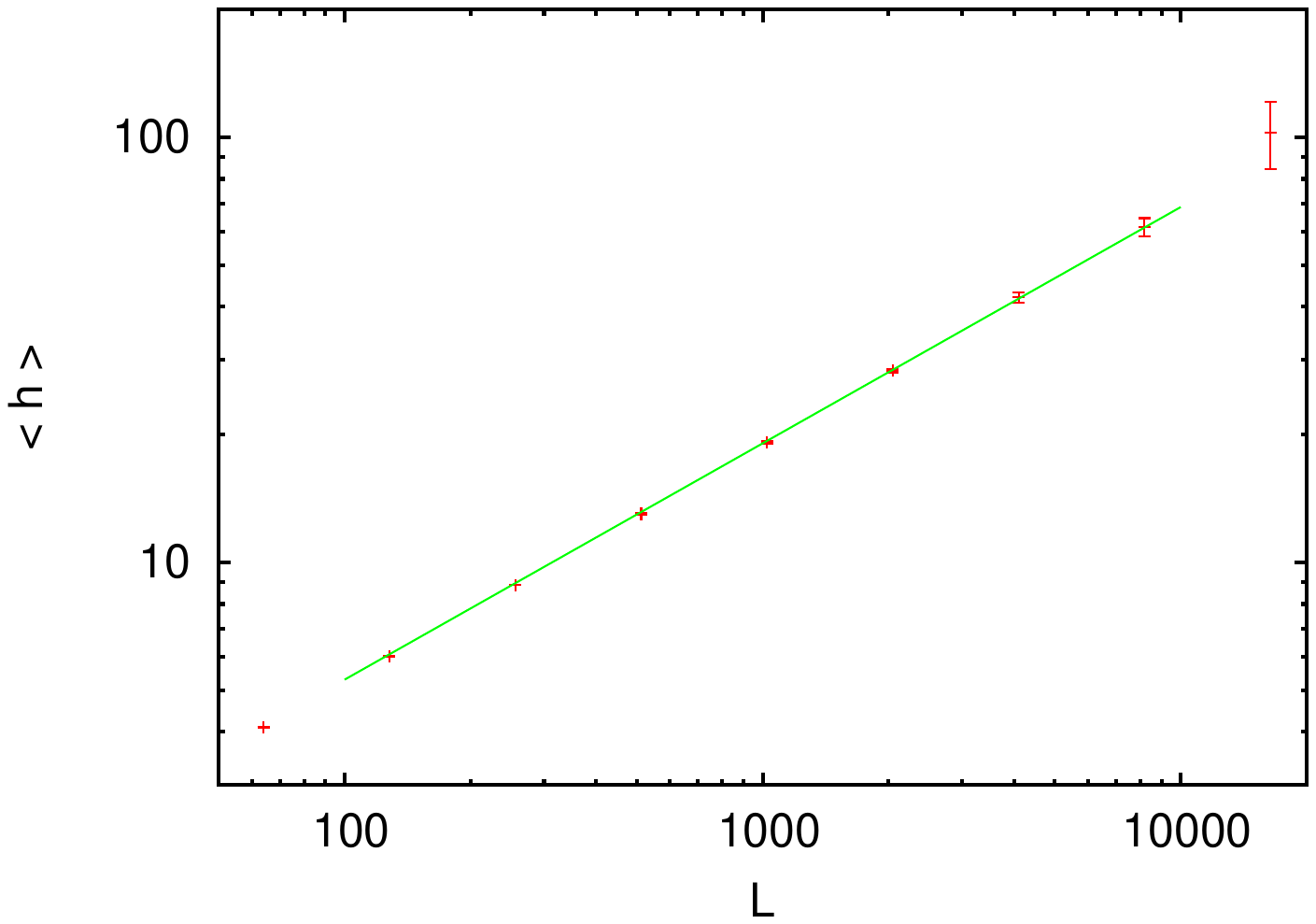}
\caption{(color online) Log-log plot of average SOS surface heights at $q=q_c$ for interdependent 
   networks on square lattices with random dependency links. The straight line indicates a power
   law with exponent 0.556.}
\label{fig.26}
\end{figure}

The dependence of the average SOS surface heights on $q$ and on $N=L^2$ is also very similar to 
that for ER networks. As functions of $q$ they show peaks that become narrower and higher with
increasing $L$, and the peak values (more precise, the values at $q=q_c$) show a power law 
\be
   \langle h(q_c) \rangle \sim L^{0.556(10)}.
\ee
(see Fig. 27).
This is in qualitative agreement with findings by D. Zhou (quoted in \cite{Danziger}), who obtained
however an exponent 0.44, in clear disagreement with our value 0.556. On the other hand, the scaling
$\langle h(q_c) \rangle \sim N^{0.278(5)}$ is the same as that for ER networks (see Eq.~(6)).

\subsubsection{Intermediate-Range Dependency Links}

Irrespective of all details, the results of the last sub-subsection show unambiguously that the 
transition is first order in the limit of infinitely long dependency links and nearest neighbor 
connectivity links, and the transition occurs roughly at $q\leq 0.7$. This should be remembered
if we now consider, following \cite{Li-Bashan,Berezin2}, dependency links with intermediate range.

In \cite{Li-Bashan,Berezin2} the dependency links were random vectors ${\bf x}=(x,y)$ with $||{\bf x}|| < r$
in the maximum norm. it was found that the model stays second order, as long as $r<8$. At $r = r^* \approx 8$
the transition becomes first order, and for $r>8$ there are claimed to be two first order transitions with
a metastable phase in between. The percolation is again implemented as site percolation, i.e. in one 
of the two lattices only a fraction $q$ of nodes is open, while all connectivity bonds and all sites
in the other lattice are open. 

\begin{figure}[htp]
\includegraphics[scale=0.31]{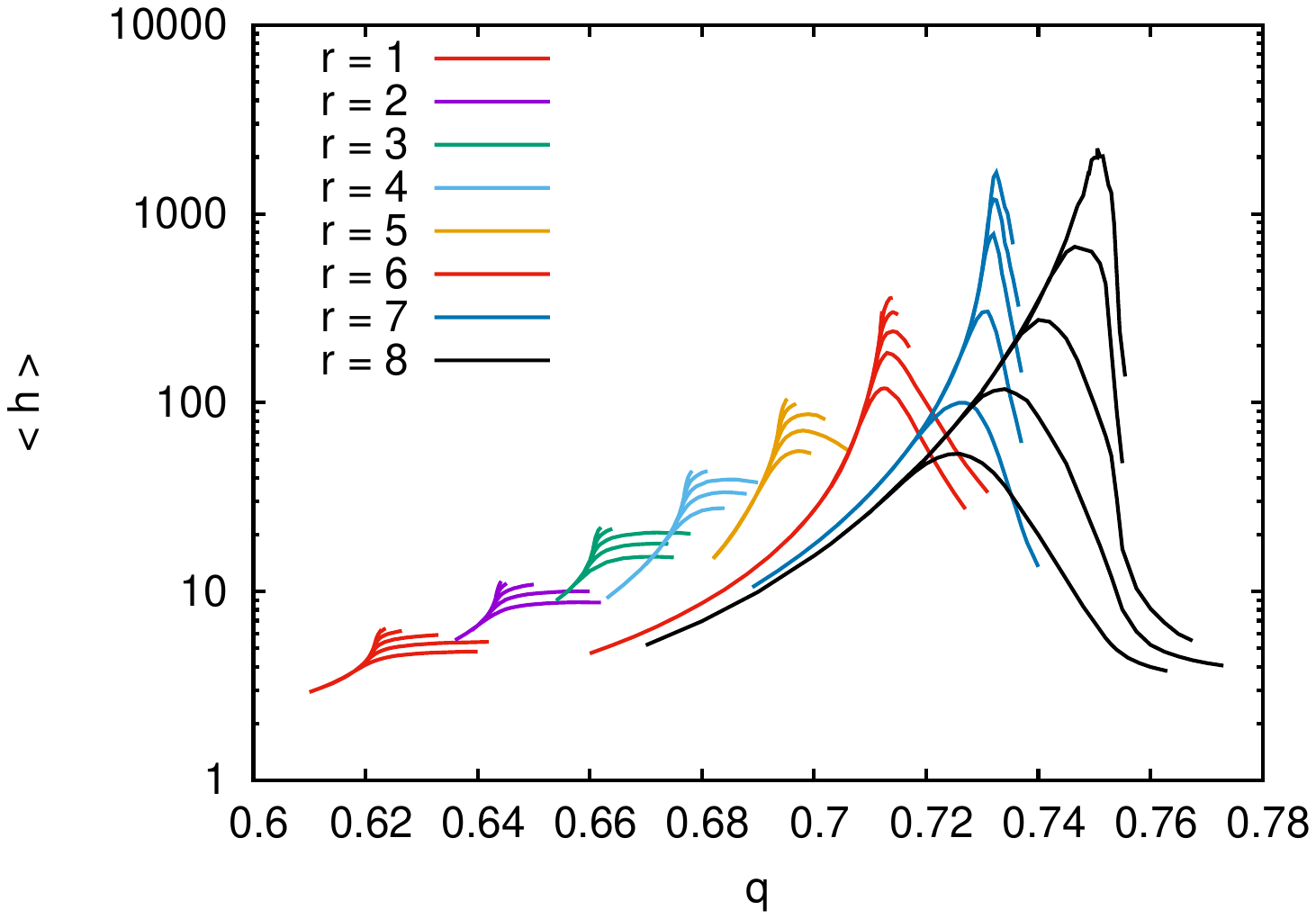}
\caption{(color online) Average SOS surface heights plotted against $q$ for interdependent 
   networks on square lattices with dependency links of lengths $\leq r$, with $r=1,2,\ldots 8$. Each group of 
   curves corresponds to the same value of $r$ (increasing from left to right), while each curve within a 
   group corresponds to a different $L$ (with $L=512, 1024,\ldots 8192$). For each $r\neq 7$ the critical 
   point is at the position where the curves for large $L$ are steepest. Our estimate for $r^*$ is slightly
   larger than $r^* = 7$.}
\label{fig.27}
\end{figure}

We verified that there is an abrupt change of the transition type at $r=r^*$, although we found $r^*$ somewhat 
closer to 7 than to 8, see Fig.~28. The most conspicuous feature in the range $1 \leq r < r^*$
is a dramatic increase of the surface heights. For all $r$ the critical points are at those values of $q$ where 
$\langle h\rangle$ changes fastest, which is to the left of the peaks for $r < 7$ and to its right for $r=8$.
For all $r < 7$ we found that the transition is not in the OP
universality class, but in the universality class discussed in subsection III.B. This is most clearly 
seen for $r=1$, where simulations are fastest (due to the smallness of $\langle h\rangle$) and where 
also corrections to scaling are smallest. One might have anticipated that there are large scaling corrections 
for $r=1$ because of the nearness to $r=0$ where the model would be just ordinary site percolation, but this 
is not true. Indeed, we found critical exponents in perfect agreement with the estimates of subsection III.B,
and with even somewhat smaller statistical errors. We do not show any plots as they are so similar to those
in subsection III.B.

For $r>r^*$ the transitions {\it look} very much like first order transitions, in agreement with the claims 
in \cite{Li-Bashan,Berezin2,Danziger}. If so, then the transition should be tricritical exactly 
at $r^*$, but we were not able to measure tricritical exponents. 
As discussed in \cite{Li-Bashan,Berezin2,Danziger},
what happens at $r=r^*$ is that fronts become unstable. Consider a scenario where open sites
exist only in half space $x>0$, while all sites with $x<0$ are closed. In ordinary percolation this
would correspond just to a boundary where all clusters are cut off at $x<0$. This case of 
percolation in the presence of boundaries has been studied in detail \cite{Grass92,Cardy92}. In
particular, the density of the giant cluster near the boundary is reduced and is governed by a 
new critical index, but otherwise the boundary has no effect on the bulk behavior. 

This is still true for $r<r^*$, but not for $r>r^*$. Although we disagree with 
\cite{Li-Bashan,Berezin2,Danziger} on details and on the theoretical treatment, we agree
with them that nodes near the boundary are removed from the viable cluster by dependencies, 
which leads then to more removals, etc., so that such boundaries create moving fronts which 
finally destroy the entire viable cluster -- unless $q$ is large and $r$ is sufficiently 
close to $r^*$. If the front established in this latter case has a sharp profile and the density 
behind it is finite (which is clearly suggested by the simulations), one has a first order transition. 

\begin{figure}[htp]
\includegraphics[scale=0.31]{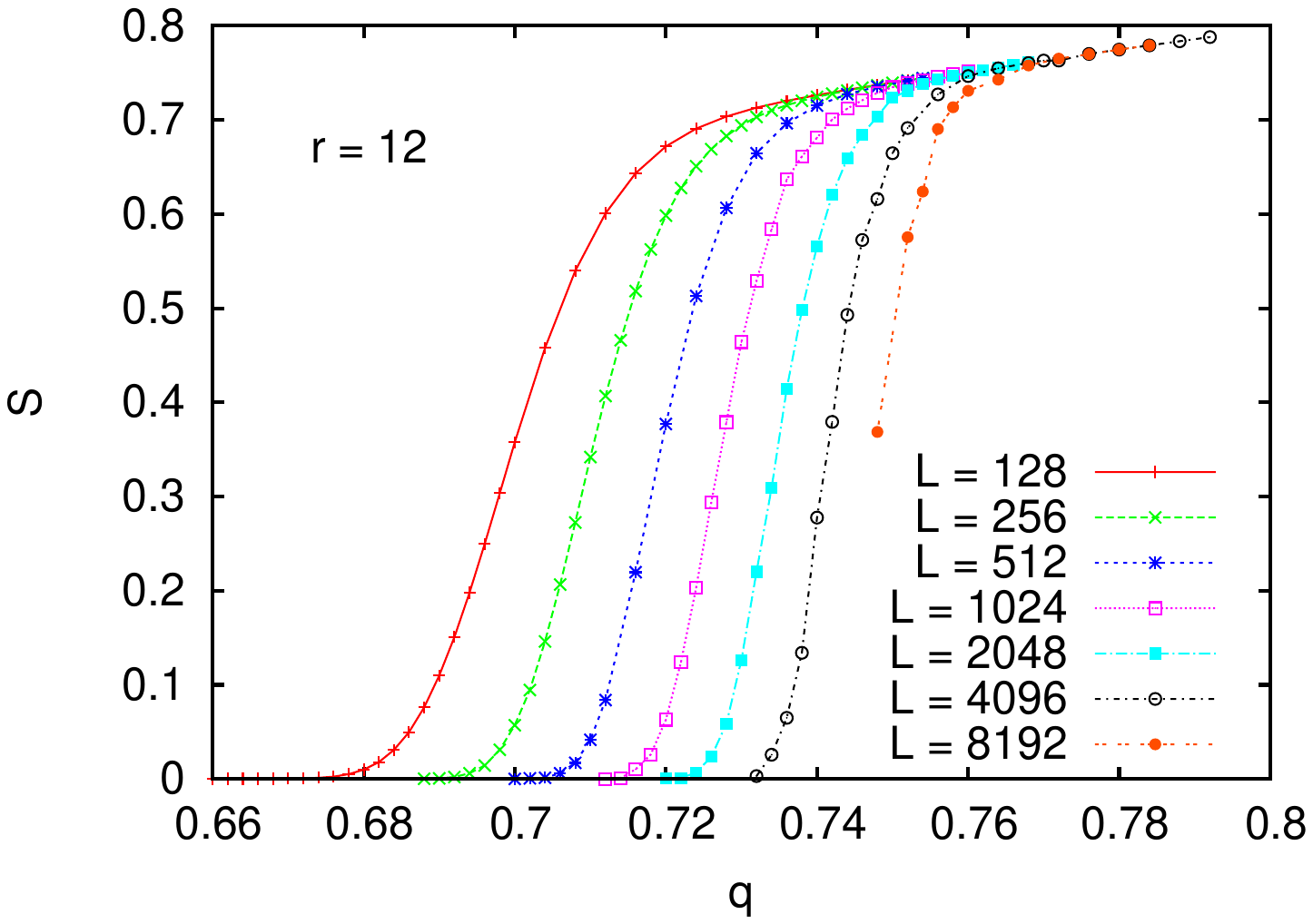}
\caption{(color online) Densities of the largest viable cluster for the model of \cite{Li-Bashan}
   with $r = 12$, plotted against $q$ for various system sizes. Here, averages are taken over 
   all runs, including those with no giant viable cluster.}
\label{fig.28}
\end{figure}

This gives thus indeed rise to the ``upper" of the two claimed transitions for $r>r^*$ (where 
$q_c(r) > q_c(r^*) \approx 0.74$) \cite{Li-Bashan,Berezin2,Danziger} -- with an important caveat 
discussed below. But we found no indication 
for the lower transition, where $q_c$ is supposed to be lower than the critical value at $r^*$.
In Fig.~29 we show the order parameter for $r=12$ (i.e., well above $r^*$) for system sizes between
$L=128$ and $8192$. According to \cite{Li-Bashan,Danziger}, there should be a first order phase 
transition at $q=0.74$. We see that there is a rather rapid cross-over at $q=0.74$, if $L \approx 2000$
(which is presumably the size on which the claim of \cite{Li-Bashan,Danziger} was based), but we 
can definitely rule out a real phase transition within the range $0.68 < q < 0765$. Indeed,
extrapolating to larger $L$ we can be rather sure that no transition occurs for any $q < 0.8$. 
Notice that $q_c = 0.753(1)$ for $r=8$ (data not shown), so the transition point is definitely 
increasing with increasing $r$. We cannot say numerically whether it agrees with the transition
defined via the propagation of fronts (is at $q\approx 0.83$ for $r=12$), but 
this seems to be the only plausible option.

Analogous results were obtained for $r=20$ and $r=30$ (data not shown). Again we find rapid but 
nevertheless smooth crossovers at values of $q$ which increase rapidly with $L$. In both
cases the cross-over happens at the ``transition points" seen in \cite{Li-Bashan,Danziger} when
$L \approx 10^3$, but we can rule out real phase transitions anywhere near these points.

\begin{figure}[htp]
\includegraphics[scale=0.31]{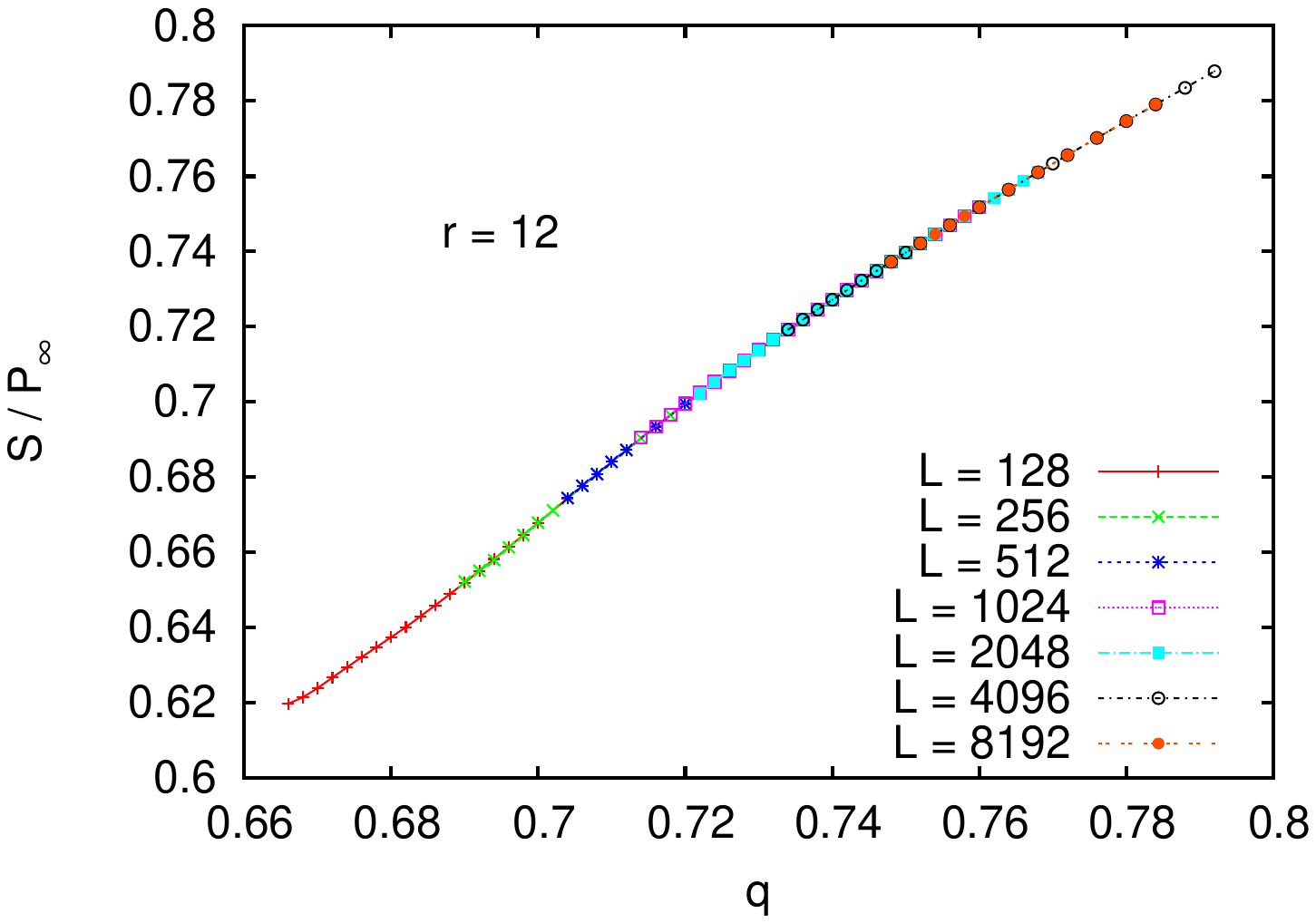}
\caption{(color online) Densities of the largest viable cluster for the model of \cite{Li-Bashan}
   with $r = 12$ (same data as in Fig.~28), but conditioned on clusters which do have a giant 
   viable cluster.}
\label{fig.29}
\end{figure}

\begin{figure}[htp]
\includegraphics[scale=0.31]{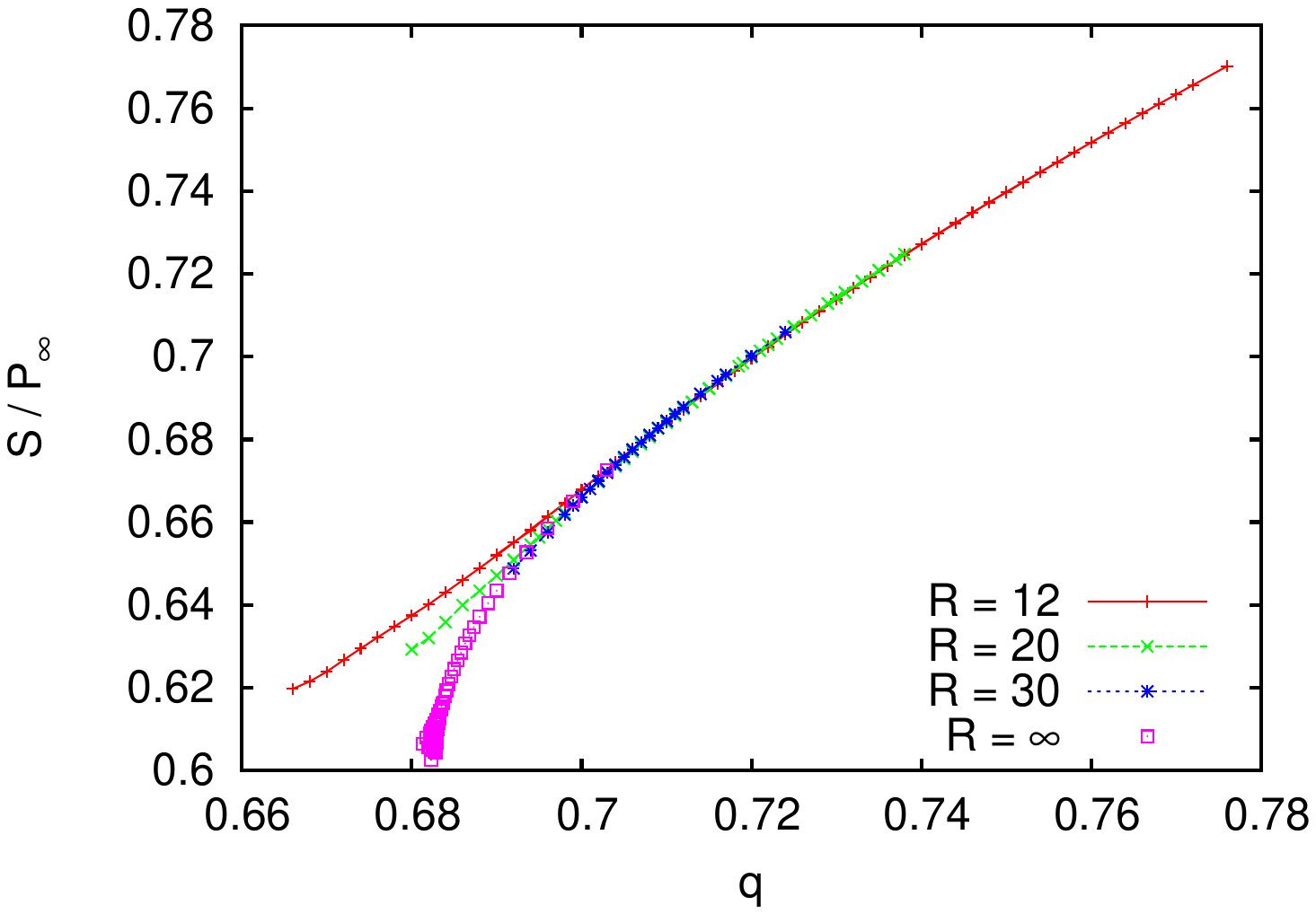}
\caption{(color online) Densities of giant viable clusters for $r = 12, 20, 30$, and $\infty$.}
\label{fig.30}
\end{figure}

Further insight is obtained by plotting not the order parameter averaged over all runs, but by
conditioning onto runs with a giant cluster. Again it is very easy to distinguish between runs with
and without giant clusters (although the distinction is not as sharp as for $r=\infty$). The 
results for $r=12$ are plotted in Fig.~30. The perfect data collapse shows that the difference 
between the curves for different $L$ seen in Fig.~29 is {\it entirely} due to differences in 
the probabilities $P_L$ with which a giant cluster is reached. The structures of the 
giant clusters themselves are completely independent of $L$. In Fig.~31, these data are 
plotted together with analogous
results for $r=20$ and $r=30$, and with the results for $r=\infty$ (obtained in the previous 
sub-subsection; see the inset in Fig.~26) extrapolated to $L\to\infty$. All four curves seem to 
become identical for large $q$, while they fan out systematically for small $q$. This figure
suggests that there would be very little dependence on $r$ (except for very small $q$), if there 
were not a mechanism which would dramatically affect the probability for a giant viable cluster
to occur.

\begin{figure}[htp]
\includegraphics[scale=0.31]{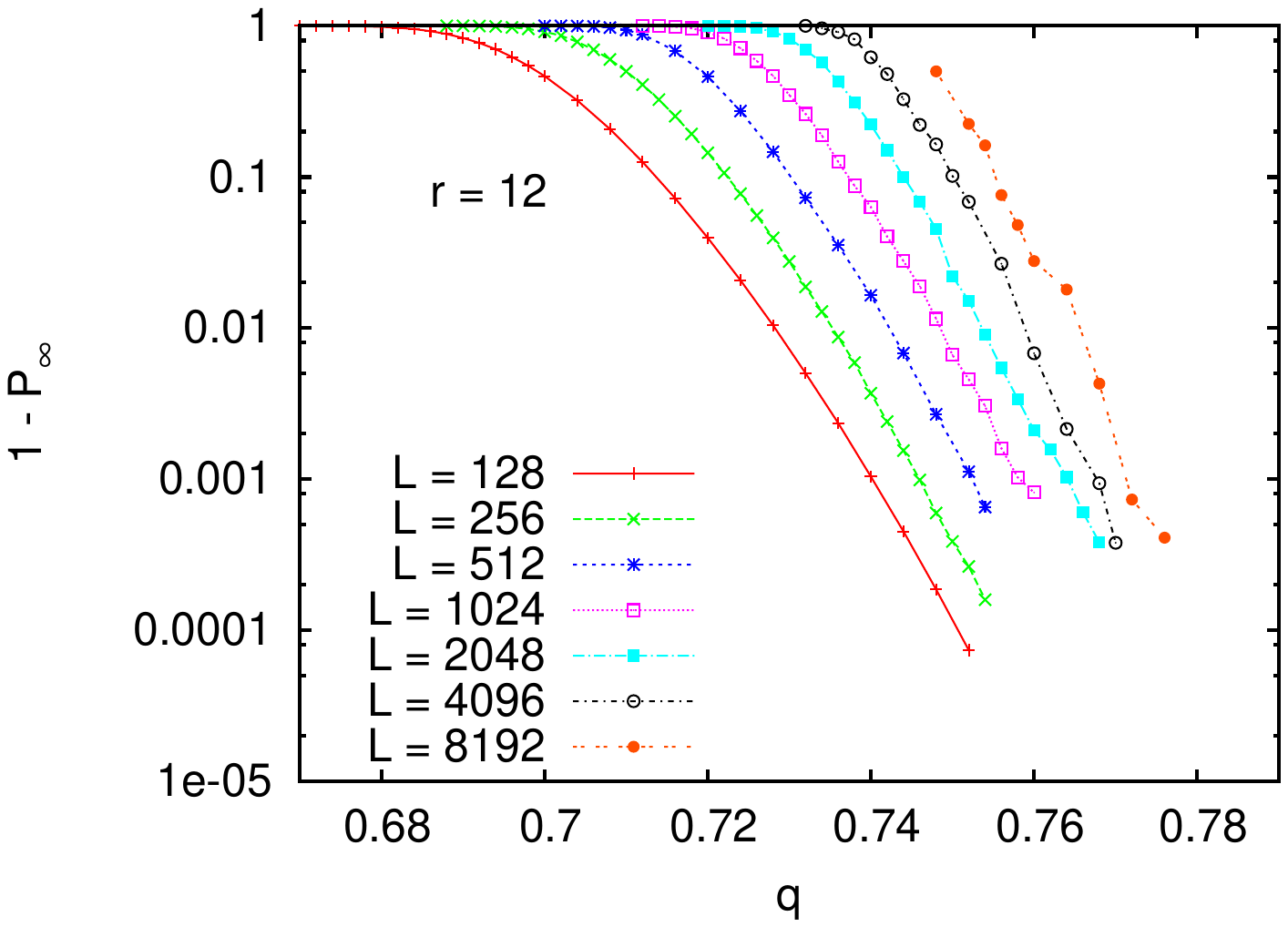}
\caption{(color online) Probabilities $1-P_\infty$ that {\it no} giant viable cluster is formed,
for $r=12$.}
\label{fig.31}
\end{figure}

This mechanism is, as suggested in \cite{Li-Bashan,Berezin2,Danziger}, the existence of voids in
the initial configuration and their subsequent growth. This is also confirmed by plotting the 
probabilities that a giant viable cluster is {\it not} formed, see Fig.~32 for $r=12$.
For large $q$ (i.e., if these probabilities are small), they increase with $L$ as 
\be
   1-P_\infty \sim c(q) L^2,
\ee
suggesting that giant cluster formation is prevented by a rare {\it extensive} mechanism, i.e. by
a mechanism whose probability to arise is $\propto L^2$. This is of course true for the formation
of voids in the initial configuration. We also see that these probabilities decrease, for fixed 
$L$, roughly exponentially with $q$. This is again what we would expect for the 
random formation of voids, with
\be
   p(m) \sim \frac{L^2}{m}(1-q)^m
\ee
being roughly the probability to form a void of $m$ sites. Numerical estimates of void sizes 
based on this estimate and on Fig.~32 give void sizes which are typically half as large as those
quotes in \cite{Li-Bashan}. This is not unreasonable, given the fact that voids were assumed in 
\cite{Li-Bashan} to be circular disks, while they could indeed have different shapes.

Thus we do agree with basic features proposed in \cite{Li-Bashan,Berezin2,Danziger}, but we 
do not agree with the existence of a second phase transition curve. We also would not call
``metastable" the states for $q$ slightly above this supposed transition curve, since 
metastability applies to systems subject to stochastic (not frozen) noise, and refers to states
which are actually unstable on long time scales. In the present case we would rather speak
of ``conditional stability", since the viable clusters are absolutely stable, but they arise
only conditioned on particular initial configurations.

\subsubsection{A Final Remark on First Order Transitions and an Alternative Scenario}

As we said, the data {\it suggest} for $r>r^*$ a first order transition, and the location of 
this transition {\it seems} to be on the curve given in \cite{Li-Bashan,Berezin2,Danziger}. 
Although we have no direct numerical evidence to doubt this, there is indirect evidence and 
theoretical arguments. As suggested in \cite{Li-Bashan,Berezin2,Danziger}, the transition 
for $r>r^*$ is essentially a percolation transition, where the spreading of voids induced by 
dependencies becomes critical. In $d\geq 3$ there exist indeed tricritical points in generalized
percolation models where the phase transition changes from second to first order \cite{Janssen,Bizhani2}.
There is however strong evidence \cite{Bizhani2,Dahmen,Bizhani3} that such tricritical points
are absent in $d=2$. This seems related to the Aizenman-Wehr theorem \cite{Aizenman2} on absence of 
first order phase transitions in random 2-d systems, although details are far from clear. 
Physically, it corresponds to the fact that phase boundaries cannot stay flat in 1+1 - dimensional 
isotropic random systems, but become crumpled on large scales. This is e.g. what is found in the 
$T=0$ random field Ising model \cite{Dahmen,Bizhani3}, and such interfaces seem always to be in 
the OP universality class, even if they look very smooth on small scales  \cite{Note3}. 

\begin{figure}[htp]
\includegraphics[scale=0.31]{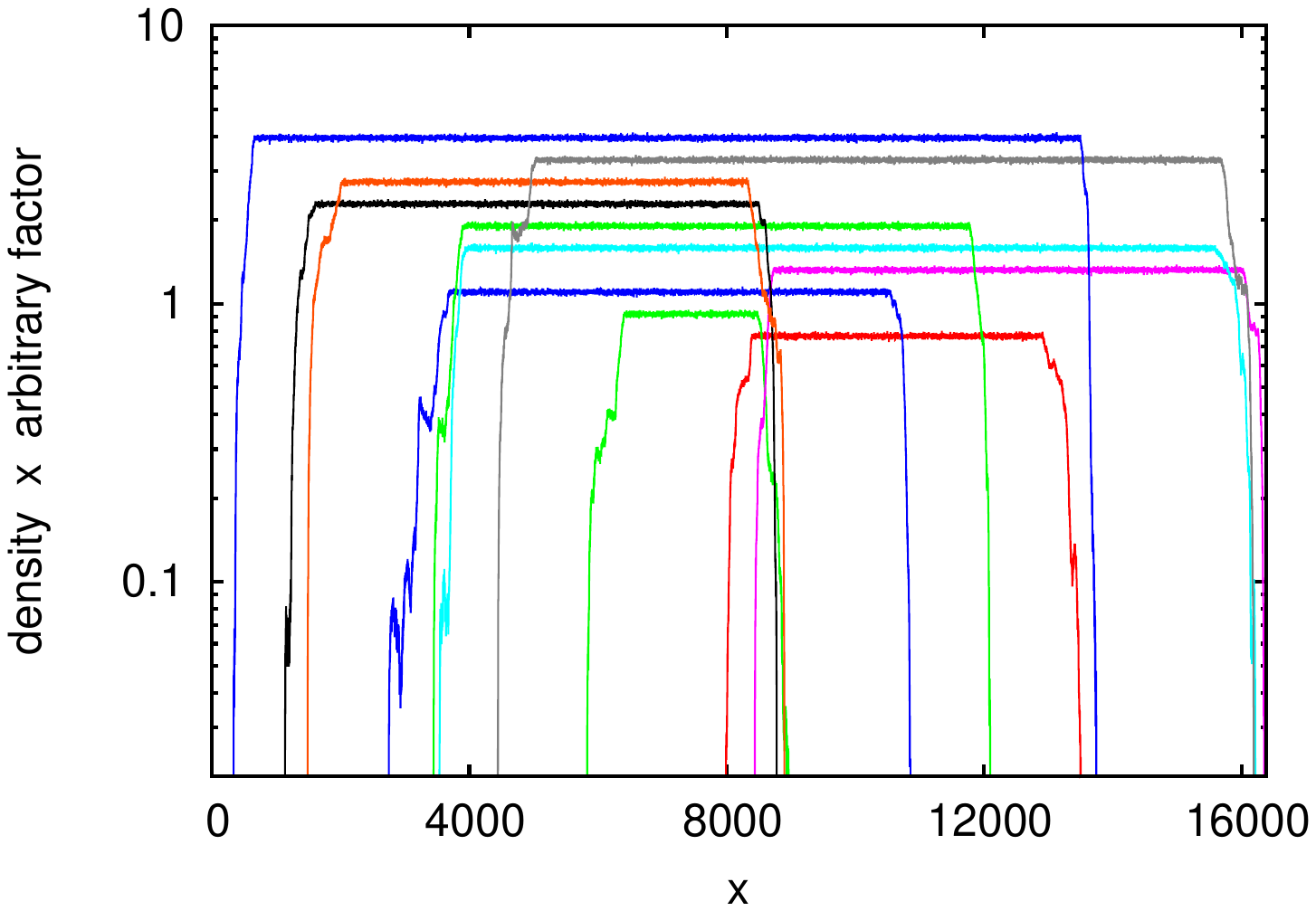}
\caption{(color online) Densities of giant viable clusters on 2-d lattices of size
   $L_\perp \times L_\| = 2048\times 16384$, with $r=9$ and $q = 0.772$ (which is our best estimate for 
   $q_c$ at $r=9$). Shown are the 1-d projections of these densities for ten typical realizations, 
   multiplied with arbitrary factors to avoid overlaps. Due to local isotropy, we would expect that 
   interface widths should become $\sim L_\perp$ in the large system limit. Actual interfaces are much 
   sharper.}
\label{fig.32}
\end{figure}

As suggested by Fig.~33, this is not true for simulations on lattices of size $L_\perp \times L_\|$ with
$L_\| \gg L_\perp$, and with zero initial densities in the regions $x<x_- = 20$ and $x>x_+ = L_\|-20$. The latter enforce 
interfaces which would stay near $x_\pm$ if $q>q_c$, but move inwards
for $q<q_c$. We find, even for $r$ very close to $r^*$,
that interfaces at $q\approx q_c$ are very straight, with fluctuations much smaller than $L_\perp$. 
For larger values of $r$ (data not shown) the effect was much enhanced, and the interfaces were essentially 
straight lines.
This is clearly a finite size effect, which is particularly enhanced by the specific choice of dependency
links made in \cite{Li-Bashan,Berezin2,Danziger}: By choosing such links to be in a square region with 
sharp boundaries, fluctuations of the interface are strongly suppressed. 

We thus conclude that all simulations of \cite{Li-Bashan,Berezin2,Danziger} are far from the true
asymptotic region. It might be that the positions of the critical points are nevertheless correct, but
they may also be far off. In any case, the fact that the transitions for $r>r^*$ {\it look} like first
order cannot be taken as evidence that they really are so. Explosive percolation \cite{Achlioptas} should
be a warning that not all transitions that {\it look} first order also are so. In the present case, 
the length scale introduced by $r$ should be enough to confuse the picture.

Indeed, we conjecture that the transition is in the universality class of OP not for $r<r^*$ (as claimed 
in \cite{Li-Bashan}), but for $r>r^*$ -- where it just corresponds to ordinary percolation of voids.
We have to admit that we do not have direct numerical evidence to support this claim, nor do we see
any possibility to obtain evidence in favor or against it in the near future.

\section{Discussion}

The purpose of this paper was twofold: On the one hand, we proposed an efficient algorithm to 
simulate viable clusters in multiplex networks, and the percolation transitions related to them.
This was motivated by the fact that exact results for this problem are obtainable only in
mean field theory (i.e. on random locally loopless graphs), and there exist ample indications 
that such predictions can be very misleading in real applications. Indeed, the algorithm presented
in this paper is not only fast, but by mapping the problem onto a solid-on-solid model we were 
able also to discuss some non-trivial structures which might become important for themselves.

The other aim of the paper was to apply this algorithm in several simple networks and lattices,
in order to test previous claims and to understand better the nature of the percolation
transition(s) in this model. We found several surprises:

\begin{itemize}
\item While the geometric structure of viable clusters on ER networks is indeed as predicted by mean 
field theory, it seems that the (pseudo-)dynamics of cascades is not. We speculated that this 
might be because the clusters are supercritical during the cascades, but more work is needed 
to understand this.
\item On finite-dimensional lattices the model is not in the universality class of ordinary
percolation, but represents its own new universality class. This is seen most dramatically in four and 
five dimensions, where OP would show continuous transitions with non-trivial anomalous exponents.
Rather, we found a first order transition in $d=5$, and a continuous transition with $\beta=1$
in $d=4$. Thus it seems as if the model has upper critical dimension $d_u=4$, although the true
mean field solution has a discontinuous transition. In $d=2$ the results are less dramatic, but 
it seems clear that it is not in the OP universality class. In $d=3$, finally, we cannot make
a clear statement because of very strong corrections to scaling.
\item While the order parameter exponent is $\beta=1$ for $d=4$ as in other models at the upper 
critical dimension, other exponents like the fractal cluster dimension and the correlation length 
are different. They seem to be simple rational numbers, but precise estimates are difficult due
to strong (logarithmic?) corrections.
\item In general, all first order transitions are hybrid. Thus, the order parameter makes on
ER networks and on 5-dimensional lattices not only a jump, but it also shows a power law. More 
surprisingly, in $d=5$ there seems also to be a non-trivial scaling mass distribution of finite
clusters.
\end{itemize}

As regards geometric networks embedded in low dimensions with semi-local links, we partly
confirmed the scenario found in \cite{Li-Bashan}. We do not find the double phase transitions
claimed there 
when the lengths of these links are are above a (tri-)critical value. We do find that there exists 
such a (tri-)critical value above which the transition {\it seems} to be first order, but we 
give strong evidence that this is related to very large finite size corrections.
More importantly, 
we claim that the model studied in \cite{Li-Bashan,Berezin2,Danziger} is very unrealistic
in assuming dependency links to be much longer than connectivity links. Whether a model where 
this relation is reversed shows similar behavior is an open question.

Other open questions concern the behavior of multiplex networks with $>2$ types of links. We have 
not yet tried to extend our algorithm to this case. Even less clear is the behavior in case of
non-mutual (asymmetric) dependencies, which cannot be mapped onto multiplex networks at all. 
The same is true for directed multiplex networks.
Finally, the true asymptotic behavior in the 2-d case with long range dependency links could,
eventually, only be solved analytically or by means of a different model realization which avoids
the introduction of a new length scale.

\section*{Acknowledgments}

Part of this work was done when I was visiting the Institute for Advanced Studies in the 
Basic Sciences in Zanjan, Iran. I am indebted for its great hospitality. For discussions, I
want to thank in particular Drs. Nahid Azimi, Ehsan Nedaee Oskoee, and Walter Nadler. For 
correspondence I am indebted to Nu\~no Ara\'ujo and Seung-Woo Son. I also want to thank the 
Complexity Science Group at the University of Calgary for their very generous granting of 
computer time.


\begin{thebibliography}{}

\bibitem{Araujo} N.A.M. Ara\'ujo, P. Grassberger, B. Kahng, K.J. Schrenk, and R.M. Ziff, 
         European Phys. J. {\bf 223}, 2307 (2014).
\bibitem{Callaway} D.S. Callaway, J.E. Hopcroft, J.M. Kleinberg, M.E.J. Newman, and S.H. Strogatz, 
          Phys. Rev. E {\bf 64}, 041902 (2001).
\bibitem{Singh} S. Boettcher, V. Singh, and R.M. Ziff, Nature Commun. {\bf 3}, 787 (2012).
\bibitem{Bizhani} G. Bizhani, P. Grassberger, and M. Paczuski, Phys. Rev. E {\bf 84}, 066111 (2011).
\bibitem{Lau} H.W. Lau, M. Paczuski, and P. Grassberger, Phys. Rev. E {\bf 86}, 011118 (2012).
\bibitem{Achlioptas} D. Achlioptas, R.M. D'Souza, and J. Spencer, Science {\bf 323}, 1454 (2009).
\bibitem{Buldyrev} S.V. Buldyrev, R. Parshani, G. Paul, H.E. Stanley, and S. Havlin, Nature {\bf 464}, 1065, (2010).
\bibitem{Dodds} P.S. Dodds and D.J. Watts, Phys. Rev. Lett. {\bf 92}, 218701 (2004).
\bibitem{Janssen} H.K. Janssen, M. M\"uller, and O. Stenull, Phys. Rev. E {\bf 70}, 026114 (2004).
\bibitem{Bizhani2} G. Bizhani, M. Paczuski, and P. Grassberger, Phys. Rev. E {\bf 86}, 011128 (2012).
\bibitem{Chen} L. Chen, F. Ghanbarnejad, W. Cai, and P. Grassberger, Europhys. Lett. {\bf 104}, 50001 (2013).
\bibitem{Cai} W. Cai, F. Ghanbarnejad, L. Chen, and P. Grassberger, to be published (2014).
\bibitem{Aizenman} M. Aizenman and C.M. Newman, Commun. Math. Phys. {\bf 107}, 611 (1986).
\bibitem{Grass2013} P. Grassberger, J. Stat. Mech. P04004 (2013).
\bibitem{Chalupa} J. Chalupa, P.L. Leath, and G.R. Reich, J. Phys. C: Solid State Phys. {\bf 12} L31 (1979).
\bibitem{Adler} J. Adler, Physica A {\bf 171}, 453 (1991).
\bibitem{Goltsev} A. V. Goltsev, S. N. Dorogovtsev, and J. F. F. Mendes, Phys. Rev. E {\bf 73}, 056101 (2006).
\bibitem{Note1} A forerunner of bootstrap percolation was the threshold model of
             Granovetter \cite{Granovetter}.
\bibitem{Baxter} G.J. Baxter, S.N. Dorogovtsev, A.V. Goltsev, J.F.F. Mendes, Phys. Rev. E {\bf 83}, 051134 (2011). 
\bibitem{Grass2011} P. Grassberger, C. Christensen, G. Bizhani, S.-W. Son, and M. Paczuski,
      Phys. Rev. Lett. {\bf 106}, 225701 (2011).
\bibitem{NSW} M. E. J. Newman, S.H. Strogatz, and D.J. Watts, Phys. Rev. E {\bf 64}, 026118 (2001).
\bibitem{Newman} M. E. J. Newman, Phys. Rev. E {\bf 66}, 016128 (2002).
\bibitem{Bollobas} B. Bollob\'as, {\it Random Graphs} (Springer, New York 1998).
\bibitem{Son2} S.-W. Son, G. Bizhani, C. Christensen, P. Grassberger, and M. Paczuski, Europhys. Lett. {\bf 97}, 16006 (2012).
\bibitem{Parshani} R. Parshani, S.V. Buldyrev, and S. Havlin, Phys. Rev. Lett. {\bf 105}, 048701 (2010).
\bibitem{Gao} J. Gao, S.V. Buldyrev, H.E. Stanley, and S. Havlin, Nature Physics {\bf 8}, 40 (2012).
\bibitem{Son1} S.-W. Son, P. Grassberger, and M. Paczuski, Phys. Rev. Lett. {\bf 107}, 195702 (2011).
\bibitem{Karrer-a} B. Karrer and M.E.J. Newman, Phys. Rev. E {\bf 82}, 016101 (2010).
\bibitem{Karrer-b} B. Karrer, M.E.J. Newman, and L. Zdeborov\'a, e-print arXiv 1405.0483 (2014).
\bibitem{Shrestha} M. Shrestha and C. Moore, e-print arXiv:1312.2070 (2013).
\bibitem{Shiraki} Y. Shiraki and Y. Kabashima, Phys. Rev. E {\bf 82}, 036101 (2010).
\bibitem{Watanabe} S. Watanabe and Y. Kabashima, e-print arXIV:1308.1210 (2013).
\bibitem{Bashan1} A. Bashan and S. Havlin, J. Stat. Phys. {\bf 145}, 686 (2011).
\bibitem{Bashan} A. Bashan, R. Parshani, and S. Havlin, Phys. Rev. E {\bf 83}, 051127  (2011).
\bibitem{Li-Bashan} W. Li, A. Bashan, S.V. Buldyrev, H.E. Stanley, and S. Havlin, Phys. Rev. Lett. {\bf 108}, 228702 (2012).
\bibitem{Berezin1} A. Bashan, Y. Berezin, S.V. Buldyrev, and S. Havlin, Nature Physics {\bf 9}, 667 (2013).
\bibitem{Berezin2} Y. Berezin, A. Bashan, M.M. Danziger, D. Li, and S. Havlin, e-print arXiv:1310.0996 (2013).
\bibitem{Baxter-b} G.J. Baxter, S.N. Dorogovtsev, A.V. Goltsev, J.F.F. Mendes, Phys. Rev. Lett. {\bf 109}, 248701 (2012).  
\bibitem{Valdez} L.D. Valdez, P.A. Macri, H.E. Stanley, and L.A. Braunstein, Phys. Rev. E {\bf 88}, 050803 (2013).
\bibitem{Cellai} D. Cellai, E. L\'opez, J. Zhou, J.P. Gleeson, and G. Bianconi, e-print arXiv:1307.6359 (2013).
\bibitem{Stippinger} M. Stippinger and J. Kert\'esz, e-print arXiv:1312.1993 (2013).
\bibitem{Danziger} M.M. Danziger, A. Bashan, Y. Berezin, and S. Havlin, J. Complex Networks, to be published (2014).
\bibitem{Zhou} D. Zhou, A. Bashan, R. Cohen, Y. Berezin, N. Shnerb, and S. Havlin, arXiv:1211.2330v3 (2014).
\bibitem{Hwang} S. Hwang, S. Choi, Deokjae Lee, and B. Kahng, e-print arXiv:1409.1147 (2014).
\bibitem{Note2} There exists also a fast 
      algorithm to find the largest viable cluster \cite{Schneider}, but since only the largest cluster is 
      followed during the cascade, it might give wrong results in cases where an initially large cluster is taken 
      over during the cascade by a cluster that had started smaller. This does not happen on random tree-like graphs,
      but it might happen on more general graphs. 
\bibitem{Privman} V. Privman and N. $\rm \check{S}$vraki\'c, J. Stat. Phys. {\bf 51}, 1111 (1988).
\bibitem{Leath} P.L. Leath, Phys. Rev. B {\bf 14}, 5046 (1976).
\bibitem{Min-Goh} B. Min and K.-I. Goh, e-print arXiv:1401.1587 (2014).
\bibitem{Note5} Notice that our families of waves are similar but yet different from the ``cascades" discussed 
     in \cite{Buldyrev}. The latter are nested sequences of clusters which all contain the maximal viable cluster, 
     while in our algorithm the nested clusters all contain the same seed $i$.
\bibitem{Dhar} D. Dhar, Phys. Rev. Lett. {\bf 64}, 1613 (1990).
\bibitem{Berezin3} Y. Berezin, A. Bashan, and S. Havlin, Phys. Rev. Lett. {\bf 111}, 189601 (2013).
\bibitem{Son3} S.W. Son, P. Grassberger, and M. Paczuski, Phys. Rev. Lett. {\bf 111}, 189602 (2013).
\bibitem{Lorenz} C.D. Lorenz and R.M. Ziff, Phys. Rev. E {\bf 57}, 230 (1998).
\bibitem{Deng} Y.J. Deng and H.W.J. Blote, Phys. Rev. E {\bf 72}, 016126 (2005).
\bibitem{Grass2013b} P. Grassberger, J. Stat. Phys. {\bf  153}, 289 (2013).
\bibitem{Grass92} P. Grassberger, J. Phys. A: Mathematical and General {\bf 25}, 5867 (1992).
\bibitem{Cardy92} J.L. Cardy, J. Phys. A: Mathematical and General {\bf 25}, L201 (1992).
\bibitem{Dahmen} B. Drossel and K. Dahmen, European Phys. J. {\bf 3}, 485 (1998).
\bibitem{Bizhani3} G. Bizhani, M. Paczuski, and P. Grassberger, to be published.
\bibitem{Aizenman2} M. Aizenman and J. Wehr, Phys. Rev. Lett. {\bf 62}, 2503 (1989).
\bibitem{Note3} Notice that interfaces in 1+1-dimensional systems are smooth, as long as they are not pinned but moving, 
      as in the KPZ problem or Eden growth.
\bibitem{Granovetter} M. Granovetter, Am. J. Sociol. {\bf83}, 1420 (1978).
\bibitem{Schneider} C.M. Schneider, N.A.M. Ara\'ujo, and H.J. Herrmann, Phys. Rev. E {\bf 87}, 043302 (2013).
\bibitem{Note4} Because sites are indexed for helical b.c. by single indices, this leads to slightly simpler 
     and faster codes. For the sizes used in this paper, periodic and helical b.c. lead to numerically indistinguishable 
     results. 


\end{thebibliography}
\end{document}